\documentclass[prb,twocolumn,amsmath,aps]{revtex4}
\usepackage{graphicx,color}
\usepackage{bm}

\def\tsigma{\tilde\sigma}

\begin{document}

\title{Wave-function approach to Master equations for quantum transport and measurement}

\author{Shmuel Gurvitz}
\email{shmuel.gurvitz@weizmann.ac.il}
\affiliation{Department of Particle Physics and Astrophysics\\
Weizmann Institute of Science, Rehovot 76100, Israel}

\date{}


\begin{abstract}
This paper presents a comprehensive review of the wave-function approach for
derivation of the number-resolved Master equations, used for description of  transport and measurement in mesoscopic systems. The review contains important amendments, clarifying subtle points in derivation of the Master equations and their validity. This completes the earlier works on the subject. It is demonstrated that the derivation does not assume weak coupling with the environment and reservoirs, but needs only high bias condition. This condition is very essential for validity of the Markovian Master equations, widely used for a phenomenological description of different physical processes.
\end{abstract}

\maketitle
\section{Introduction}
Quantum transport through mesoscopic systems is
one of the most extensively investigated areas of theoretical
physics \cite{data}. Typically, the transport takes place when a mesoscopic system, characterized by isolated energy levels, is connected with reservoirs, which energy spectrum is continuous (very dense). In addition, the system is interacting with the environment, like with phonon (photon) reservoirs with continuous energy spectrum. As a results, the transport trough mesoscopic systems would display an interplay of quantum and classical features.

For evaluation of current through the mesoscopic system (or other related quantities), one needs to trace over infinite number of the environmental and reservoirs states that are not directly observed. Such a procedure should be carried out in the density matrix of an {\em entire} system, $\rho(s,s',I,I',t)$, where $s(s')$ and $I(I')$ denote all variables of the mesoscopic system and the environment and reservoirs, respectively. As a result we obtain the reduced density matrix of the mesocopic system, $\sigma (s,s',t)={\rm Tr}_I^{}[\rho(s,s',I,I,t))]$ that  determines its behavior and bears all effects of the environment and reservoirs. Our goal, therefore is to derive the Master equations for the reduced density-matrix $\sigma(s,s',t)$.

In the case of transport through single quantum dot, phenomenological ``classical'' rate equations has been used long ago \cite{glaz,ak,been,glp,Davies4603}. These included only the diagonal density-matrix elements. The situation becomes different for coupled wells (dots) with aligned levels. The quantum transport through these devices goes on via quantum superposition between the states in adjacent wells. As a result, non-diagonal density matrix elements should appear in the equations of motion. These terms have no classical counterparts, and therefore the classical rate equations have to be modified. A plausible phenomenological modification of master equations for some particular cases of the resonance tunneling through double-dot structures has been proposed in \cite{naz,glp1} by using an analogy to the optical Bloch equations\cite{bloch}.

Since then the Master equations approach to quantum transport has been extensively developed \cite{gp,2,weimin,xinqi}. However, most of the derivations were based on the second order perturbative expansion of the total density matrix, $\rho$. As an alternative, we proposed a different approach based on derivation of the Master equations, from the Schr\"odinger equation for the total many-body wave-function, $i\partial_t|\Psi (t)\rangle=H|\Psi (t)\rangle$ \cite{gp,g,g1}. In such  framework, all approximations are applied to the wave-function. The latter is  used to build up the density matrix, which finally is traced over the environmental and reservoir states. This technique allow us to obtain the Master equations of motion for different physical problems in a most simple way. The resulting equation is obtained in a form of generalized Lidndblad \cite{lind} equation, which in addition  describes a back-action on the environment. This would allow us to use this equation to study the counting statistics and the continuous measurement process.

This paper is a comprehensive review of the original papers Refs.~[\onlinecite{gp,g,g1,eg2,xq2,g2}], with some amendments and corrections. These are  elaborating subtle points of derivations, which were not properly addressed in earlier publications. The paper is organized as follows. Sec.~II-VI present detailed derivation of Master equations for different mesoscopic structures. Sec.~VII presents extension of the previous results to general case of multi-dot systems. Sec. VIII-IX consider application of the Master equations approach to a description of the continuous measurement process, without an explicit use of the Projection postulate. Sec.~X deals with continuous measurement of tunneling to non-Markovian reservoirs in a relation with the quantum Zeno effect. Last Section contains concluding remarks.

\section{Single-well structure}

\begin{figure}[tbh]
\includegraphics[width=7cm]{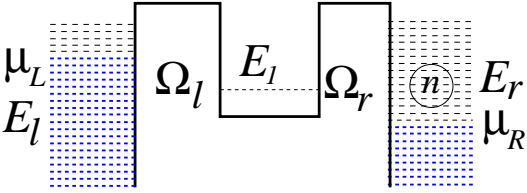}
\caption{Resonant tunneling through a single-level dot (see text), $n$ denoted number of electrons arriving the right reservoir at time $t$.}.
\label{fig1}
\end{figure}

Let us consider a mesoscopic ``device'' consisting of a quantum
well (dot), coupled to two separate electron reservoirs.
The density of states in the reservoirs is very high (continuum).
The dot, however, contains only isolated levels.
We first demonstrate how to achieve the reduction of many-body
Schroedinger equation to the rate equation in the simplest example,
Fig.~\ref{fig1}, with only one level, $E_1$, inside the dot.
We also ignore the Coulomb electron-electron interaction inside the well and the spin degrees of freedom. Hence, only one electron may occupy the well.

With the stand simplifications, the tunneling Hamiltonian of the entire
system in the occupation number representation is
\begin{align}
&H =\sum_l E_{l}a^{\dagger}_{l}a_{l} +
E_1 a_1^{\dagger}a_{1} +\sum_r E_{r}a^{\dagger}_{r}a_{r}\nonumber\\
&+ \sum_l \Omega_{l}(a^{\dagger}_{l}a_1 +a^{\dagger}_{1}a_{l})
+ \sum_r \Omega_{r}(a^{\dagger}_{r}a_1 +a^{\dagger}_{1}a_{r})\;.
\label{Ham}
\end{align}
where $a_{1}^\dagger$ and $a_{l(r)}^\dagger$ are the creation
operators for the electron in the dot and in the left (right) reservoirs.
The subscripts $l$ and $r$ enumerate correspondingly the levels in the left (emitter) and right (collector) reservoirs. The tunneling couplings of the dots to the leads, $\Omega_{l,r}$, are weakly dependent of energy (wide-band limit) and can be chosen to be real in the absence of magnetic field.

For simplicity,  we restrict ourselves to the zero temperature in the reservoirs. All the levels in the emitter and the
collector are initially filled with electrons up to
the Fermi energy $\mu_L$ and $\mu_R$, respectively.
This situation will be treated as the ``vacuum'' state $|0\rangle$.

This vacuum state is unstable; the Hamiltonian
Eq.~(\ref{Ham}) requires it to decay exponentially to a continuum state
having the form $a_{1}^{\dagger}a_{l}|0\rangle$ with an electron in
the level $E_1$ and a hole in the emitter continuum. These continuum
states are also unstable and decay to states
$a_{r}^{\dagger}a_{l} |0\rangle$
having a particle in the
collector continuum as well as a hole in the emitter continuum, and
no electron in the level $E_1$. The latter, in turn, are decaying into
the states $a_{1}^{\dagger}a_{r}^{\dagger}a_{l}a_{l'} |0\rangle$
and so on. The evolution of the whole system is described by the
many-particle wave function, which is represented as
\begin{align}
|\Psi (t)\rangle &= \left [ b_0(t) + \sum_l b_{1l}(t)
           a_{1}^{\dagger}a_{l} + \sum_{l,r} b_{lr}(t)
           a_{r}^{\dagger}a_{l}\right.\nonumber\\
&\left.+ \sum_{l<l',r} b_{1ll'r}(t)
           a_{1}^{\dagger}a_{r}^{\dagger}a_{l}a_{l'}
           + \ldots \right ] |0\rangle,
\label{wf}
\end{align}
where $b(t)$ are the time-dependent probability amplitudes to
find the system in the corresponding states described above with the initial
condition $b_0(0)=1$, and all the other $b(0)$'s being zeros.
Substituting Eq.~(\ref{wf}) into in the
Schr\"odinger equation $i|\dot\Psi (t)\rangle =H|\Psi (t)\rangle$,
results in an infinite set of coupled linear differential equations for the
amplitudes $b(t)$. Applying the Laplace transform
\begin{equation}
\tilde{b}(E)=\int_0^{\infty}e^{iEt}b(t)dt
\label{lap}
\end{equation}
and taking account of the initial conditions, we transform
the linear differential
equations for $b(t)$ into an infinite set of algebraic equations
for the amplitudes $\tilde b(E)$,
\begin{subequations}
\label{ineq}
\begin{align}
& E \tilde{b}_{0}(E) - \sum_l \Omega_{l}\tilde{b}_{1l}(E)=i
\label{ineq1}\\
&(E + E_{l} - E_1) \tilde{b}_{1l}(E) - \Omega_{l}
      \tilde{b}_{0}(E) - \sum_r \Omega_{r}\tilde{b}_{lr}(E)=0
\label{ineq2}\\
&(E + E_{l} - E_{r}) \tilde{b}_{lr}(E) -
      \Omega_{r} \tilde{b}_{1l}(E) -
      \sum_{l'} \Omega_{l'}\tilde{b}_{1ll'r}(E)=0
\label{ineq3}\\
&(E + E_{l} + E_{l'} - E_1 - E_{r}) \tilde{b}_{1ll'r}(E)-
\Omega_{l'} \tilde{b}_{lr}(E)\nonumber\\
&~~~~~~~~~~~~~~~~~~+\Omega_{l} \tilde{b}_{l'r}(E)-
\sum_{r'} \Omega_{r'}\tilde{b}_{ll'rr'}(E)=0
\label{ineq4}\\
&~~~~~~~~~~~~~~~~~~~~~~~~~~~~~~\cdot
\nonumber
\end{align}
\end{subequations}
Note that in the Laplace transform $E\to E+i\delta$, where $E$ plays a role of energy in the scattering theory.

Equations (\ref{ineq}) can be substantially simplified. Let us replace
the amplitude $\tilde b$ in the term $\sum\Omega\tilde b$
of each of the equations  by
its expression obtained from the subsequent equation. For example,
substitute $\tilde{b}_{1l}(E)$ from Eq.~(\ref{ineq2}) into Eq.~(\ref{ineq1}).
We obtain
\begin{align}
&\left [ E - \sum_l \frac{\Omega^2_{l}}{E + E_{l} - E_1}
    \right ] \tilde{b}_{0}(E)\nonumber\\[5pt]
&~~~~~~~~~~~~~~~~~- \sum_{l,r}
    \frac{\Omega_{l}\Omega_{r}}{E + E_{l} - E_1}
    \tilde{b}_{lr}(E)=i.
\label{exam}
\end{align}
Since the states in the reservoirs are very dense (continuum),
one can replace the sums over $l$ and $r$ by integrals, for instance
$\sum_{l}\;\rightarrow\;\int \rho_{L}(E_{l})\,dE_{l}\:$,
where $\rho_{L}(E_{l})$ is the density of states in the emitter. We assume that it is independent of the $E_l$ (wide band limit), and the same for the density of state of the collector, $\rho_{L,R}(E_{l},E_r)\equiv \rho_{L,R}$. Also the tunneling couplings are considered as the energy independent, $\Omega_{l,r}\equiv\Omega_{L,R}(E_l,E_r)\equiv \Omega_{L,R}$. Then the first sum in Eq.~(\ref{exam}) becomes an
integral
\begin{align}
  S_1&=\int_{-\Lambda}^{\mu_L}\rho_LdE_l {\Omega_L^2\over
    E+E_l-E_1}\nonumber\\
   &=-i{\Gamma_L\over2}\Theta (\mu_L+E-E_1)+{\Gamma_L\over2\pi}\ln{\mu_L+E-E_1\over \Lambda -E+E_1}
\label{exam1}
\end{align}
where $\Lambda$ is a cut-off and $\Gamma_L=2\pi |\Omega_L|^2\rho_L$ is a partial width of the level $E_1$ due to its coupling to the emitter. (The corresponding partial width due to tunneling to the collector is
$\Gamma_R = 2\pi\rho_R(E_1)|\Omega_R(E_1)|^2$).

Now we consider the case of $|\mu_{L,R}-E_1|\gg \Gamma_{L,R}$, corresponding to the energy level $E_1$ is deeply inside the bias, $V=\mu_L-\mu_R$, (large bias limit). Then the last term of Eq.~(\ref{exam1}) vanishes in the limit of $\mu_L,\Lambda\to\infty$ as $\Gamma_{L}/(\mu_L-E_1)$ (or  $(\Gamma_{L}/(\Lambda-E_1)$). We then obtain
\begin{align}
  S_1=-i{\Gamma_L\over 2}\, .
\label{exam2}
\end{align}

Consider now the second sum in Eq.~(\ref{exam}).
\begin{align}
  S_2=\int_{\mu_R}^\Lambda\rho_RdE_r\int_{-\Lambda}^{\mu_L}\rho_L
    dE_l{\Omega_L\Omega_R
    \tilde b_{lr}(E)\over E+E_l-E_1}
\label{exam3}
\end{align}
In contrast with the first
term of Eq.~(\ref{exam}), the amplitude $\tilde b$ is not factorized
out the integral (\ref{exam3}). We refer to this type of terms as
``cross-terms''. Fortunately, all ``cross-terms'' vanish in the
limit of large bias. This greatly simplifies the
problem and is very crucial for a transformation of the
Schr\"odinger to the Master equations. The reason is that the poles of
the integrand in the $E_l(E_r)$-variable in the ``cross-terms'' are
on the same side of the integration contour. One can find it by using a
perturbation series the amplitudes $\tilde b$ in powers of $\Omega$.
For instance, from iterations of Eqs.~(\ref{ineq}) one finds
\begin{align}
  \tilde b_{lr}(E)={i\Omega_L\Omega_R
    \over E(E+E_l-E_r)(E+E_l-E_1)}+\cdots
 \label{eexam4}
\end{align}

The higher order powers of $\Omega$ have the same structure. Since
$E\to E+i\epsilon$ in the Laplace transform, all poles of the
amplitude $\tilde b_{lr}(E)\equiv b_{lr}(E,E_l,E_r)$ in the $E_l$-variable are below the
real axis. In this case, substituting Eq.~(\ref{eexam4}) into
Eq.~(\ref{exam3}) we find in the limit $\mu_L,\Lambda\to\infty$,
\begin{align}
 \int\limits_{-\infty}^{\infty}
    {\Omega_L\Omega_RdE_l
    \over (E+i\epsilon )( E+E_l-E_1+i\epsilon )^2( E+E_l-E_r+i\epsilon )}
    = 0
    \label{eexam5}
\end{align}
Thus, $S_2\to 0$ in the large bias limit.

Applying analogous considerations to the other equations of the
system (\ref{ineq}), we finally arrive at the following set of
equations:
\begin{subequations}
\label{fineq}
\begin{eqnarray}
&& (E + i\Gamma_L/2) \tilde{b}(E)=i
\label{fineq1}\\
&& (E + E_{l} - E_1 + i\Gamma_R/2) \tilde{b}_{1l}(E)
      \nonumber\\
&&~~~~~~~~~~~~~~~~~~~~~~~~~~~~~~~- \Omega_{L} \tilde{b}(E)=0
\label{fineq2}\\
&& (E + E_{l} - E_{r} + i\Gamma_L/2) \tilde{b}_{lr}(E)\nonumber\\
&& ~~~~~~~~~~~~~~~~~~~~~~~~~~~~~~-
      \Omega_{R} \tilde{b}_{1l}(E)=0
\label{fineq3}\\
&& (E + E_{l} + E_{l'} - E_1 - E_{r} + i\Gamma_R/2)
       \tilde{b}_{1ll'r}(E)\nonumber\\[5pt]
&&~~~~~~~~~~~~~~~ -\Omega_{L} \tilde{b}_{lr}(E)+\Omega_{L}
\tilde{b}_{l'r}(E)=0
\label{fineq4}\\[5pt]
& &~~~~~~~~~~~~~~~~~~~~~~~~\cdots \nonumber
\end{eqnarray}
\end{subequations}

Now we introduce the (reduced) density matrix of the ``device''.
The Fock space of the quantum well consists of only two possible states,
namely: $|a\rangle$ -- the level $E_1$ is empty,
and $|b\rangle$ -- the level $E_1$ is occupied. In this basis,
the diagonal elements of the density matrix of the ``device'',
$\sigma_{aa}$ and $\sigma_{bb}$, give the probabilities of the
resonant level being empty or occupied, respectively. In our
notation, these probabilities are represented as follows:
\begin{subequations}
\label{sigmas}
\begin{align}
\sigma_{aa} &= |b_{0}(t)|^2 + \sum_{l,r} |b_{lr}(t)|^2
            + \sum_{l<l',r<r'} |b_{ll'rr'}(t)|^2 + \ldots
\nonumber\\[5pt]
&\equiv \sigma_{aa}^{(0)} + \sigma_{aa}^{(1)}
            + \sigma_{aa}^{(2)} + \ldots\;,
\label{sigmaa}\\[5pt]
\sigma_{bb} &= \sum_l |b_{1l}(t)|^2 +
             \sum_{l<l',r} |b_{1ll'r}(t)|^2\nonumber\\
             &~~~~~~~~~~~~~~~~~~~~~~~
             + \sum_{l<l'<l'',r<r'} |b_{1ll'l''rr'}(t)|^2 + \ldots\nonumber\\[5pt]
             &
\equiv \sigma_{bb}^{(0)} + \sigma_{bb}^{(1)}
             + \sigma_{bb}^{(2)} + \ldots,
\label{sigmbb}
\end{align}
\end{subequations}
where the index $n$ in $\sigma^{(n)}$ denotes the number of electrons
in the collector, Fig.~\ref{fig1}. The current  $I(t)$ flowing through the system is
$I(t)=e\dot N_R(t)$, where $N_R(t)$ is the number of
electrons accumulated in the collector, i.e.
\begin{equation}
N_R(t) = \sum_{n} n\left [\sigma_{aa}^{(n)}(t)+\sigma_{bb}^{(n)}(t)\right ]
\label{char}
\end{equation}
In the following we adopt the units where the electron charge $e$ is unity.

The density submatrix elements are directly related
to the amplitudes $\tilde b(E)$ through the inverse Laplace transform,
\begin{equation}
\sigma^{(n,n')}(t)=
\sum_{l\ldots , r\ldots}
\int \frac{dEdE'}{4\pi^2}\tilde b_{l\cdots r\cdots}(E)
\tilde b^*_{l\cdots r\cdots}(E')e^{i(E'-E)t}
\label{invlap}
\end{equation}
By means of this equation one can transform  Eqs. (\ref{fineq})) for
the amplitudes $b(E)$ into differential equations directly to the
density matrix $\sigma_{jj'}^{(n,n')}(t)$, where $j=a,b$ denote the
state of the SET with an unoccupied or occupied dot and $n$ denotes
the number of electrons which have arrived at the collector by time
$t$, Fig.~\ref{fig1}. In fact, as follows from our derivation, the diagonal density-matrix elements, $j=j'$ and $n=n'$, form a closed system in
the case of resonant tunneling through one level. The
off-diagonal elements, $j\not =j'$, appear in the equation of motion
whenever more than one discrete level of the system carry the
transport (see in following). Therefore we concentrate below on the
diagonal density-matrix elements only,
$\sigma_{aa}^{(n)}(t)\equiv\sigma_{aa}^{(n,n)}(t)$ and
$\sigma_{bb}^{(n)}(t)\equiv\sigma_{bb}^{(n,n)}(t)$. Applying the
inverse Laplace transform on finds
\begin{widetext}
\begin{subequations}
\label{invlapp}
\begin{eqnarray}
\sigma_{aa}^{(n)}(t)&=& \sum_{l\ldots , r\ldots}\int
\frac{dEdE'}{4\pi^2}\tilde b_{\underbrace{l\cdots}_{n}
  \underbrace {r\cdots}_{n}}(E)
\tilde b^*_{\underbrace{l\cdots}_{n}
  \underbrace {r\cdots}_{n}}(E')e^{i(E'-E)t}\label{invlapa}\\
\sigma_{bb}^{(n)}(t)&=& \sum_{l\ldots , r\ldots}\int
\frac{dEdE'}{4\pi^2}\tilde b_{1{\underbrace{l\cdots}_{n+1}
  \underbrace {r\cdots}_{n}}}(E)
\tilde b^*_{1{\underbrace{l\cdots}_{n+1}
  \underbrace {r\cdots}_{n}}}(E')e^{i(E'-E)t}\label{invlapb}
\end{eqnarray}
\end{subequations}
\end{widetext}
Consider, for instance, the term $\sigma_{bb}^{(0)}(t)=\sum_l
|b_{1l}(t)|^2$. Multiplying Eq.~(\ref{fineq2}) by
$\tilde{b}^*_{1l}(E')$ and then subtracting the complex conjugated
equation with the interchange $E\leftrightarrow E'$ we obtain
\begin{eqnarray}
&&\int\frac{dEdE'}{4\pi^2}\sum_l\left[(E'-E -i\Gamma_R)\tilde
b_{1l}(E)\tilde b^*_{1l}(E')\right.\nonumber\\&&\left.~~-2{\mbox
{Im}}\sum_L\Omega_L \tilde b_{1l}(E)\tilde
b_0^*(E')\right]e^{i(E'-E)t}=0 \label{ap3}
\end{eqnarray}

Using Eq.~(\ref{invlapb}) one easily finds that the first integral
in Eq.~(\ref{ap3}) equals to $-i[\dot\sigma_{bb}^{(0)}(t)+
\Gamma_R\sigma_{bb}^{(0)}(t)]$. Next, substituting
\begin{align}
\tilde{b}_{1l}(E)=\frac{\Omega_{L} \tilde{b}_0(E)} {E + E_{l} -
E_1 + i\Gamma_R/2} \label{ap2}
\end{align}
from Eq.~(\ref{fineq2}) into the second term of Eq.~(\ref{ap3}), and
replacing a sum by an integral, one can perform the
$E_l$-integration in the large bias limit, $\mu_L\to\infty$,
$\mu_R\to -\infty$. Then using again Eq.~(\ref{invlapb}) one reduces
the second term of Eq.~(\ref{ap3}) to
$i\Gamma_L\sigma_{aa}^{(0)}(t)$. Finally, Eq.~(\ref{ap3}) reads
$\dot{\sigma}^{(0)}_{bb}(t)=\Gamma_L \sigma^{(0)}_{aa}(t) - \Gamma_R
\sigma_{bb}^{(0)}(t)$.

The same algebra can be applied for all other amplitudes $\tilde
b_\alpha (t)$. For instance, by using Eq.~(\ref{invlapa}) one easily finds
that Eq.~(\ref{fineq3}) is converted to the following rate equation
$\dot{\sigma}^{(1)}_{00}(t)=-\Gamma_L \sigma^{(1)}_{aa}(t) +\Gamma_R
\sigma_{bb}^{(0)}(t)$. With respect to the states involving more
than one electron (hole) in the reservoirs (the amplitudes like
$\tilde b_{1ll'r}(E)$ and so on), the corresponding equations
contain the Pauli exchange terms. By converting these equations into
those for the density matrix using our procedure, one finds the
``cross terms'', like $\sum\Omega_{L}^2 \tilde{b}_{l'r}(E)
\tilde{b}^*_{lr}(E')$, generated by Eq.~(\ref{fineq4}). Yet, these
terms vanish after an integration over $E_{l(r)}$ in the large bias
limit, as the second term in Eq.~(\ref{exam}). The rest of the algebra remains the same, so one obtains
$\dot{\sigma}^{(1)}_{bb}(t)=\Gamma_L \sigma^{(1)}_{aa}(t)
- \Gamma_R \sigma_{bb}^{(1)}(t)$.
Finally we arrive to the following infinite system of the
chain equations for the diagonal elements,
$\sigma_{aa}^{(n)}$ and $\sigma_{bb}^{(n)}$, of the density matrix,
\begin{subequations}
\label{aandb}
\begin{align}
&\dot{\sigma}^{(0)}_{aa}(t) = - \Gamma_L \sigma^{(0)}_{aa}(t)\;,
\label{anought}\\
&\dot{\sigma}^{(0)}_{bb}(t) = \Gamma_L \sigma^{(0)}_{aa}(t)
                               - \Gamma_R \sigma_{bb}^{(0)}(t)\;,
\label{bnought}\\
&\dot{\sigma}^{(1)}_{aa}(t) = - \Gamma_L \sigma^{(1)}_{aa}(t)
                               + \Gamma_R \sigma_{bb}^{(0)}(t)\;,
\label{aone}\\
&\dot{\sigma}^{(1)}_{bb}(t) = \Gamma_L \sigma^{(1)}_{aa}(t)
                               - \Gamma_R \sigma_{bb}^{(1)}(t)\;,
\label{bone}\\
&~~~~~~~~~~~~~~~~~~~\cdots
\nonumber
\end{align}
\end{subequations}
Summing up these equations, one easily obtains
differential equations for the
total probabilities $\sigma_{aa}=\sum_n \sigma_{aa}^{(n)}$ and
 $\sigma_{bb}=\sum_n \sigma_{bb}^{(n)}$:
\begin{subequations}
\label{a4}
\begin{align}
\dot\sigma_{aa} & =  -\Gamma_L\sigma_{aa}+\Gamma_R\sigma_{bb}\;,
\label{a4a}\\
\dot\sigma_{bb} & =  \Gamma_L\sigma_{aa}-\Gamma_R\sigma_{bb}\;,
\label{a4b}
\end{align}
\end{subequations}
which should be supplemented with the initial conditions
\begin{equation}
\sigma_{aa}(0)=1, \;\;\;\;\sigma_{bb}(0)=0.
\label{init}
\end{equation}

Using Eqs. (\ref{char}), (\ref{aandb}) we obtain
the total current
\begin{align}
I(t)&=\dot N_R(t)= \Gamma_R [\sigma_{bb}^{(0)}(t) + \sigma_{bb}^{(1)}(t)
       + \sigma_{bb}^{(2)}(t) + \ldots ]\nonumber\\[5pt]
        &= \Gamma_R \sigma_{bb}(t).
\label{curr}
\end{align}
Thus the current $I(t)$ is directly proportional to
the charge density in the well. Solving Eqs.~(\ref{a4})
and substituting $\sigma_{bb}(t)$ into Eq.~(\ref{curr}), we obtain (for
$t\to\infty$) the standard formula for the dc resonant
current,
\begin{equation}
I=\frac{\Gamma_L\Gamma_R}{\Gamma_L+\Gamma_R}\;.
\label{aa5}
\end{equation}
Notice that whereas the time-behavior of the current $I(t)$ depends on the
initial condition,
the stationary current $I=I(t\to\infty)$, Eq.~(\ref{aa5}), does not.

Equations (\ref{a4}), derived from the many-body Schr\"odinger equation,
coincide with the classical rate equations in the sequential picture for
the resonant tunneling. However, as shows our derivation these equations can be valid only when the resonance energy is inside the bias, and
$\Gamma_{L,R}^{}\ll |E_1-\mu_{L,R}^{}|$ (large bias limit). If the resonance is near the Fermi energy of a reservoir, the time-dependent Scr\"odinger equation cannot be reduced to the rate equations (\ref{a4}).

\section{Coulomb blockade}

Now we extend the approach of Sect.~II to include the
effects of Coulomb interaction. Consider again the
quantum well in Fig.~\ref{fig1}, taking into
account the spin degrees of freedom ($s$). In this case
the tunneling Hamiltonian~(\ref{Ham}) becomes
\begin{align}
H &= \sum_{l,s} E_{l}a^{\dagger}_{ls}a_{ls} +
              \sum_{s}E_1 a^{\dagger}_{1s}a_{1s} +
              \sum_{r,s} E_{r}a^{\dagger}_{rs}a_{rs}
\nonumber\\
          &  + \sum_{l,s} \Omega_{l}(a^{\dagger}_{ls}a_{1s} +
              a^{\dagger}_{1s}a_{ls})
              + \sum_{r,s} \Omega_{r}(a^{\dagger}_{rs}a_{1s} +
              a^{\dagger}_{1s}a_{rs})\nonumber\\
&+Ua^{\dagger}_{1s}a_{1s}a^{\dagger}_{1,-s}a_{1,-s}\; ,
\label{c1}
\end{align}
where $s=\pm 1/2$, and $U$ is the Coulomb repulsion energy.

Writing down the many-body wave function, $|\Psi (t)\rangle$, in
the occupation number representation, just as in
Eq.~(\ref{wf}), and then substituting it into
the Schr\"odinger equation $i|\dot\Psi (t)\rangle =H|\Psi (t)\rangle$,
we find a system of coupled equations
for the amplitudes $b(t)$
\begin{subequations}
\label{c2}
\begin{align}
&E \tilde{b}_{0}(E) - \sum_l \Omega_{l}\left [\tilde{b}_{\uparrow l}(E)
+\tilde{b}_{\downarrow l}(E)\right ]=i
\label{c2a}\\
&(E + E_{l} - E_1) \tilde{b}_{\uparrow l}(E) - \Omega_{l}\tilde{b}_{0}(E)
\nonumber\\
&~~~~~~~~~
-\sum_{l'} \Omega_{l'}\tilde{b}_{\uparrow \downarrow ll'}(E)-\sum_r \Omega_{r}\tilde{b}_{lr}(E)=0
\label{c2b}\\
&(E+E_l-E_r)\tilde{b}_{lr}(E)- \Omega_{r}\tilde{b}_{\uparrow l}(E)\nonumber\\
&~~~~~~~~~-
\sum_{l'} \Omega_{l'}\left [\tilde{b}_{\uparrow l'}(E)
+\tilde{b}_{\downarrow l'}(E)\right ]=0
\label{c2c}\\
&(E + E_{l} + E_{l'}- 2E_1-U) \tilde{b}_{\uparrow \downarrow ll'}(E)
      -\Omega_{l'} \tilde{b}_{\uparrow l}(E)\nonumber\\
&-\Omega_{l}\tilde{b}_{\downarrow l'}(E)
-\sum_{r}\Omega_{r} \left [\tilde{b}_{\uparrow ll'r}(E)
+\tilde{b}_{\downarrow ll'r}(E)\right ]=0
\label{c2d}\\
&~~~~~~~~~~~~~~~~~~~~~~~~~~~~~~~~\cdots
\nonumber
\end{align}
\end{subequations}
In order to shorten notations we eliminated the index (1) of the level
$E_1$ in the amplitudes $b$,
so that $\tilde b_{\uparrow(\downarrow )\ldots}(t)$
denotes the probability amplitude
to find one electron inside the well with spin up (down), and
the  amplitude $\tilde b_{\uparrow\downarrow\ldots}(t)$ is the probability
amplitude to find two electrons inside the well.

Equations~(\ref{c2}) can be simplified
by using the same procedure as described in the previous section.
For instance, by substituting  $\tilde{b}_{lr}$ from Eq.~(\ref{c2c}) and
$\tilde{b}_{\uparrow \downarrow ll'}$ from
Eq.~(\ref{c2d}) into Eq.~(\ref{c2b}), and neglecting the ``cross terms''
on the grounds of the same arguments as
in the analysis of Eq.~(\ref{exam})), we obtain
\begin{align}
&\Big[E+E_l-E-\int_{-\infty}^{\mu_L}
\frac{\rho_L(E_{l'})\Omega_L^2(E_{l'})dE_{l'}}
{E + E_{l} + E_{l'}- 2E_1-U}\nonumber\\[5pt]
&~~~~~~~~~~
-\int_{\mu_R}^{\infty}\frac{\rho_R(E_{r})\Omega_R^2(E_{r})dE_{r}}
{E+E_l-E_r}\Big]\tilde b_{\uparrow l}(E)=0
\label{cc2}
\end{align}
where $\Omega_{L,R}^{}(E_{l,r})\equiv \Omega_{l,r}^{}$. Since $E_l\sim E_1$, the singular parts of the integrals in (\ref{cc2}) are respectively
$\;\,-i\Theta (\mu_L^{}+E-E_1+U)\,\Gamma'_L/2$ and
$\;\,-i\Theta (E+E_1-\mu_R^{})\,\Gamma_R/2$, where
\begin{align}
&\Gamma_{L(R)}=2\pi\rho_{L(R)}(E_1)|\Omega_{L(R)}(E_1)|^2,\nonumber\\
&\Gamma_{L(R)}'=2\pi\rho_{L(R)}(E_1+U)|\Omega_{L(R)}(E_1+U)|^2.
\label{cc3}
\end{align}
Here $\rho_{L(R)}$ is the spin up or spin down density of states
in the emitter (collector),
$\rho_{L(R)}\equiv\rho_{L(R)\uparrow}=\rho_{L(R)\downarrow}$.
As in the previous section, we assume the resonance level being
deeply inside the bias, $|E_1-\mu_{L,R}|\gg \Gamma_{L,R}$. If,
in addition, $E_1+U\ll \mu_L^{}$, the theta-function in the singular
parts of the integrals in (\ref{cc2}) can be replaced by one.
In the opposite case, $E_1+U\gg \mu_L^{}$, the corresponding singular part
is zero.

Proceeding this way with the other equations of the
system (\ref{c2}), we finally obtain
\begin{subequations}
\label{c3}
\begin{align}
& (E +i\Gamma_L)\tilde{b}_{0}(E)=i
\label{c31}\\
&(E + E_{l} - E_1+i\Gamma_L'/2+i\Gamma_R/2) \tilde{b}_{\uparrow l}(E)\nonumber\\
&~~~~~~~~~~~~~~~~~~~~~~~~~~~~~~~~~~~~ -
\Omega_{l}\tilde{b}_{0}(E) =0
\label{c32}\\
&(E+E_l-E_r+i\Gamma_L)\tilde{b}_{lr}(E)- \Omega_{r}\tilde{b}_{\uparrow l}(E)=0
\label{c23}\\
&(E + E_{l} + E_{l'}- 2E_1-U+i\Gamma_R')
\tilde{b}_{\uparrow\downarrow ll'}(E)\nonumber\\
&~~~~~~~~~~~~~~~~~~~~
      -\Omega_{l} \tilde{b}_{\downarrow l'}(E)
+\Omega_{l'} \tilde{b}_{\uparrow l}(E)=0
\label{c34}\\
&~~~~~~~~~~~~~~~~~~~~~~~~~~~~\cdots
\nonumber
\end{align}
\end{subequations}

Equations (\ref{c3}) can be transformed
into equations for the density matrix of the ``device'' by using the method
of the previous section. Since the algebra remains essentially
the same, we give only
the final equations for the diagonal density matrix elements
$\sigma_{aa}^{(n)}(t)$, $\sigma_{bb\uparrow}^{(n)}(t)$,
$\sigma_{bb\downarrow}^{(n)}(t)$ and $\sigma_{cc}^{(n)}(t)$. These are
the probabilities to find: a) no electrons inside
the well; b) one electron with spin up (down) inside the well, and
c) two electrons inside the well, respectively.
The index $n$ denotes the number of
electrons accumulated in the collector. We obtain
\begin{subequations}
\label{c4}
\begin{align}
\dot\sigma_{aa}^{(n)} & =-2\Gamma_L\sigma_{aa}^{(n)}
+\Gamma_R\sigma_{bb\uparrow}^{(n-1)}+\Gamma_R\sigma_{bb\downarrow}^{(n-1)}
\label{c4a}\\
\dot\sigma_{bb\uparrow}^{(n)} & = -(\Gamma'_L+\Gamma_R)
\sigma_{bb\uparrow}^{(n)}+\Gamma_L\sigma_{aa}^{(n)}
+\Gamma'_R\sigma_{cc}^{(n-1)}
\label{c4b}\\
\dot\sigma_{bb\downarrow}^{(n)} & = -(\Gamma'_L+\Gamma_R)
\sigma_{bb\downarrow}^{(n)}+\Gamma_L\sigma_{aa}^{(n)}
+\Gamma'_R\sigma_{cc}^{(n-1)}
\label{c4c}\\
\dot\sigma_{cc}^{(n)} & = -2\Gamma'_R\sigma_{cc}^{(n)}
+\Gamma'_L\sigma_{bb\uparrow}^{(n)}+\Gamma'_L\sigma_{bb\downarrow}^{(n)}
\label{c4d}
\end{align}
\end{subequations}
These rate equations look as a generalization of the rate
equations (\ref{aandb}), if one allows the well to be occupied by two
electrons.
The Coulomb repulsion leads merely to a modification of the
corresponding rates $\Gamma\to\Gamma'$, due to increase of the
two electron energy.

Summing up the partial probabilities we obtain
for the total probabilities,
$\sigma(t)=\sum_n\sigma^{(n)}(t)$, the following equations:
\begin{subequations}
\label{c5}
\begin{align}
&\dot\sigma_{aa}  = -2\Gamma_L\sigma_{aa}
+\Gamma_R\sigma_{bb\uparrow}+\Gamma_R\sigma_{bb\downarrow}
\label{c5a}\\
&\dot\sigma_{bb\uparrow}  = -(\Gamma'_L+\Gamma_R)
\sigma_{bb\uparrow}+\Gamma_L\sigma_{aa}
+\Gamma'_R\sigma_{cc}
\label{c5b}\\
&\dot\sigma_{bb\downarrow}  = -(\Gamma'_L+\Gamma_R)
\sigma_{bb\downarrow}+\Gamma_L\sigma_{aa}
+\Gamma'_R\sigma_{cc}
\label{c5c}\\
&\dot\sigma_{cc}  = -2\Gamma'_R\sigma_{cc}
+\Gamma'_L\sigma_{bb\uparrow}+\Gamma'_L\sigma_{bb\downarrow},
\label{c5d}
\end{align}
\end{subequations}
It follows from Eq.~(\ref{c5}) that
$\sigma_{aa}(t)+\sigma_{bb\uparrow}(t)+\sigma_{bb\downarrow}(t)+\sigma_{cc}(t)=1$. Respectively one finds for the current
\begin{align}
&I(t)=\sum_n n[\dot\sigma^{(n)}(t)]=
\Gamma_R\left [\sigma_{bb\uparrow}(t)+\sigma_{bb\downarrow}(t)\right ]\nonumber\\
&~~~~~~~~~~~~~~~~~~~~~~~~~~~~~~~~~~~~~~~~~~~~
+2\,\Gamma'_R\sigma_{cc}(t)
\label{c6}
\end{align}

Equations (\ref{c5}), (\ref{c6}) can be solved most easily for dc
current, $I=I(t\to\infty )$. In this case $\dot\sigma =0$, and
Eqs.~(\ref{c5}) turn into the system of linear algebraic equations.
Finally we obtain
\begin{equation}
I=\frac{2\Gamma_L\Gamma'_R(\Gamma'_L+\Gamma_R)}{\Gamma_L\Gamma'_L+
2\Gamma_L\Gamma'_R+\Gamma_R\Gamma'_R}
\label{c7}
\end{equation}
so that dc current does not depend on the initial conditions.

If $E_1\ll \mu_L^{}\ll E_1+U$, one finds from Eq.~(\ref{cc2}) that
$\Gamma'_L=0$, so that the state with two electrons inside the
well is not available. In this case one obtains
from Eq.~(\ref{c7}) for the dc current
\begin{equation}
I=\frac{2\Gamma_L\Gamma_R}{2\Gamma_L+\Gamma_R}
\label{c8}
\end{equation}
It is interesting to note that this result
is different from Eq.~(\ref{aa5}), although
in both cases only one electron can occupy the well.
However, if the Coulomb repulsion effect is small, i.e.
$\Gamma'_{L,R}=\Gamma_{L,R}$, Eq.~(\ref{c7}) does produce the same result as
Eq.~(\ref{aa5}), provided the density of states is doubled
due to the spin degrees of freedom.

One can also consider the case when
the Fermi level in the right reservoir $\mu_R$ lies above the resonance
level $E_1$, but below $E_1+U$, so that $\Gamma_R=0$, Eq.~(\ref{cc2}).
Then the resonant transitions of electrons from the left to the
right reservoirs can go only through the state with two
electrons inside the well. Using Eq.~(\ref{c7}) one finds
for the dc current
\begin{equation}
I=\frac{2\Gamma'_L\Gamma'_R}{\Gamma'_L+2\Gamma'_R},
\label{c9}
\end{equation}
which coincides with the result found by Glazmann and Matveev\cite{glaz}.

\section{Double-well structure}

\subsection{Non-interacting electrons.}
\begin{figure}[tbh]
\includegraphics[width=7cm]{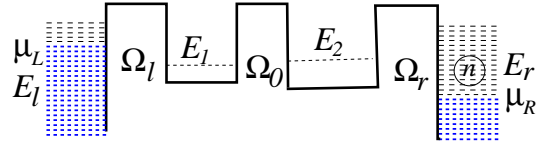}
\caption{Resonant tunneling through a double-well structure. $n$ denoted number of electrons arriving the right reservoir at time $t$.}.
\label{fig2}
\end{figure}

Now we turn to the coherent case of resonant tunneling.
Let us  consider the coupled-well structure, shown in Fig.~\ref{fig2}.
We assume that both levels $E_{1,2}$ are
inside the bias, i.e. $\mu_R\ll E_{1},E_{2}\ll \mu_L$.
In order to make our derivation as clear as possible,
we begin with the case of no spin degrees of freedom and no Coulomb
interaction.

The tunneling Hamiltonian for this system is
\begin{align}
H&= \sum_l E_{i}a^{\dagger}_{l}a_{l} +
              E_1 a_1^{\dagger}a_{1} + E_2 a_2^{\dagger}a_{2}+
              \sum_r E_{r}a^{\dagger}_{r}a_{r}
\nonumber\\
          &  +\Omega_0(a_1^{\dagger}a_{2}+a_2^{\dagger}a_{1})
              + \sum_l \Omega_{l}(a^{\dagger}_{l}a_1 +
              a^{\dagger}_{1}a_{l})\nonumber\\
&+ \sum_r \Omega_{r}(a^{\dagger}_{r}a_2 +
              a^{\dagger}_{2}a_{r})\;.
\label{b1}
\end{align}
where $a_{1,2}^{\dagger}$, $a_{1,2}$ are creation and annihilation operators
for an electron in the first or the second well, respectively. All the
other notations are taken from Sect.~II.
The many-body wave function describing this system can be written
in the occupation number representation as
\begin{align}
|\Psi (t)\rangle & =  \left [ b_0(t) + \sum_l b_{1l}(t)
     a_{1}^{\dagger}a_{l} + \sum_{l,r} b_{lr}(t)a_{r}^{\dagger}a_{l}\right.
      \nonumber\\
      &+ \left. \sum_l b_{2l}(t)a_{2}^{\dagger}a_{l}
           +\sum_{ll'} b_{12ll'}(t)a_{1}^{\dagger}a_{2}^{\dagger}a_{l}a_{l'}\right.
           \nonumber\\
           &
           \left.
           +\sum_{l<l',r} b_{1ll'r}(t) a_{1}^{\dagger}a_{r}^{\dagger}a_{l}a_{l'}
           + \ldots \right ] |0\rangle,
\label{b2}
\end{align}
Substituting Eq.~(\ref{b2}) into the Shr\"odinger equation with the
Hamiltonian (\ref{b1}) and performing the Laplace transform,
we obtain an infinite set of the coupled equations for the
amplitudes $\tilde b(t)$:
\begin{subequations}
\label{b3}
\begin{align}
& E \tilde{b}_{0}(E) - \sum_l \Omega_{l}\tilde{b}_{1l}(E)=i
\label{b3a}\\
&(E + E_{l} - E_1) \tilde{b}_{1l}(E) - \Omega_{l}
      \tilde{b}_0(E) -\Omega_{0}\tilde{b}_{2l}(E)=0
\label{b3b}\\
&(E + E_{l} - E_2) \tilde{b}_{2l}(E) -
      \Omega_0 \tilde{b}_{1l}(E) -
      \sum_{l'} \Omega_{l'}\tilde{b}_{12ll'}(E)\nonumber\\
      &~~~~~~~~~~~~~~~~~~~~~~~~~~~~~~~~~~~~-
      \sum_r \Omega_{r}\tilde{b}_{rl}(E)=0
\label{b3c}\\
&(E + E_{l} + E_{l'} - E_1 - E_2) \tilde{b}_{12ll'}(E)-
\Omega_{l'} \tilde{b}_{2l}(E)\nonumber\\
&~~~~~~~~~~~~~~~~~~+
\Omega_{l} \tilde{b}_{2l'}(E)
-\sum_{r} \Omega_{r}\tilde{b}_{1ll'r}(E)=0
\label{b3d}\\
&~~~~~~~~~~~~~~~~~~~~~~~~~~~~~~~~~~~~~~ \cdots
\nonumber
\end{align}
\end{subequations}
Using exactly the same procedure as in the previous sections, Eqs.~(\ref{exam})-(\ref{eexam5}),
we transform Eqs. (\ref{b3}) into the following set of equations:
\begin{subequations}
\label{b4}
\begin{align}
& (E + i\Gamma_L/2) \tilde{b}_{0}(E)=i
\label{b4a}\\
& (E + E_{l} - E_1) \tilde{b}_{1l}(E)
      - \Omega_{l} \tilde{b}_{0}(E)-\Omega_0\tilde b_{2l}(E)=0
\label{b4b}\\
& (E + E_{l} - E_2+i\Gamma_L/2+ i\Gamma_R/2) \tilde{b}_{2l}(E)
\nonumber\\
     &~~~~~~~~~~~~~~~~~~~~~~~~~~~~~~~~~~~~~~~
-\Omega_{0} \tilde{b}_{1l}(E)=0
\label{b4c}\\
& (E + E_{l} + E_{l'} - E_1 - E_2 + i\Gamma_R/2)
     \tilde{b}_{12ll'}(E) -
     \nonumber\\
     &~~~~~~~~~~~~~~~~~~~~~~~~~~~
       \Omega_{l'} \tilde{b}_{2l}(E) +\Omega_{l} \tilde{b}_{l'r}(E)=0
\label{b4d}\\
&~~~~~~~~~~~~~~~~~~~~~~~~~~~~~\cdots
\nonumber
\end{align}
\end{subequations}

The amplitudes $b(t)$ determine the density submatrix of
the system, $\sigma_{ij}^{(n)}$, in the corresponding Fock space:
$|a\rangle$ -- the levels $E_{1,2}$ are empty,
$|b\rangle$ -- the level $E_1$ is occupied,
$|c\rangle$ -- the level $E_2$ is occupied,
$|d\rangle$ -- the both level $E_{1,2}$ are occupied;
the index $n$ denotes the number of electrons in
the collector. The matrix elements of the density
matrix of the ``device'' can be written as
\begin{subequations}
\label{b5}
\begin{align}
&\sigma_{aa}=\sum_n\sigma_{aa}^{(n)}
\equiv |b_{0}(t)|^2 + \sum_{l,r} |b_{lr}(t)|^2\nonumber\\
&~~~~~~~~~~~~~~~~~~~~~~~~~~~~
            + \sum_{l<l',r<r'} |b_{ll'rr'}(t)|^2 + \ldots
\label{b5a}\\
&\sigma_{bb} =\sum_n\sigma_{bb}^{(n)}\equiv \sum_l |b_{1l}(t)|^2 +
             \sum_{l<l',r} |b_{1ll'r}(t)|^2\nonumber\\
&~~~~~~~~~~~~~~~~~~
+ \sum_{l<l'<l'',r<r'} |b_{1ll'l''rr'}(t)|^2 + \ldots
\label{b5b}\\
&\sigma_{cc} =\sum_n\sigma_{cc}^{(n)}\equiv \sum_l |b_{2l}(t)|^2 +
             \sum_{l<l',r} |b_{2ll'r}(t)|^2\nonumber\\
&~~~~~~~~~~~~~~~~~~
+ \sum_{l<l'<l'',r<r'} |b_{2ll'l''rr'}(t)|^2 + \ldots
\label{b5c}\\
&\sigma_{dd} =\sum_n\sigma_{dd}^{(n)}\equiv \sum_{l<l'} |b_{12ll'}(t)|^2\nonumber\\
&~~~~~~~~~+
\sum_{l<l'<l''<l''',r<r'} |b_{12ll'l''l'''rr'}(t)|^2
             + \ldots
\label{b5d}\\
&\sigma_{bc} =\sum_n\sigma_{bc}^{(n)}\equiv \sum_l b_{1l}(t)b_{2l}^*(t)\nonumber\\
&~~~~~~~~~~~~~~~~~~~~~~~~~~
+\sum_{l<l',r} b_{1ll'r}(t) b_{2ll'r}^*(t)+ \ldots
\label{b55}
\end{align}
\end{subequations}

Now we transform Eqs.~(\ref{b4}) into
differential equations for $\sigma^{(n)}(t)$.
Consider for instance the term $\sigma_{bb}^{(0)}=\sum_l |b_{1l}(t)|^2$,
Eq.~(\ref{b5b}), where the amplitudes $b_{1l}$ are determined
by Eq.~(\ref{b4b}). Multiplying Eq.~(\ref{b4b}) by $\tilde{b}^*_{1l}(E')$
and subtracting the complex conjugate equation with $E\leftrightarrow E'$,
we find
\begin{align}
&\sum_l(E'-E)\tilde b_{1l}(E)\tilde b^*_{1l}(E')\nonumber\\
&-\sum_l\Omega_l
[\tilde b^*_{0}(E')\tilde b_{1l}(E)-\tilde b_{0}(E)\tilde b^*_{1l}(E')]\nonumber\\
&
-\Omega_0\sum_l
[\tilde b^*_{2l}(E')\tilde b_{1l}(E)
-\tilde b_{2l}(E)\tilde b^*_{1l}(E')]=0
\label{b7}
\end{align}
After applying the inverse Laplace transform, Eqs.~(\ref{invlap}),(\ref{invlapp})
the first term in this equation
becomes $-i\dot\sigma_{bb}^{(0)}(t)$.
Next, substituting
\begin{equation}
\tilde{b}_{1l}(E)=\frac{
      \Omega_{l} \tilde{b}_{0}(E)+\Omega_0\tilde b_{2l}(E)}{E + E_{l} - E_1}
\label{b6}
\end{equation}
from Eq.~(\ref{b4b}) into the second term of Eq.~(\ref{b7}),
and replacing the sum by an integral over
$E_l$, we reduce this term to $i\Gamma_L\tilde b_0(E)\tilde b^*_0(E')$.
After the inverse Laplace transform it becomes
$i\Gamma_L\sigma_{aa}^{(0)}(t)$. Notice that in the large bias limit, the ``cross term'',
$\propto \Omega_0\Omega_l\tilde b_0\tilde b_{2l}$ ,
does not contribute to the integral over $E_l$, since
the poles of the integrand in the
$E_l$-variable lie on one side of the integration contour
(cf. with  Eq.~(\ref{eexam5})).
The third term of Eq.~(\ref{b7}) turns to be
$\Omega_0[\sigma_{bc}^{(0)}(t)-\sigma_{cb}^{(0)}(t)]$,
after the inverse Laplace transform. Finally we obtain a
differential equation for the density submatrix
element $\sigma_{bb}^{(0)}$,
\begin{equation}
\dot\sigma_{bb}^{(0)}(t)=\Gamma_L\sigma_{aa}^{(0)}
		+i\Omega_0(\sigma_{bc}^{(0)}-\sigma_{cb}^{(0)}).
\label{bb7}
\end{equation}
In contrast to the rate equations of the previous sections,
the diagonal matrix element $\sigma_{bb}$ is coupled with the
off-diagonal density matrix element $\sigma_{bc}$.

The corresponding differential equation for $\sigma_{bc}$ can
be easily obtained by multiplying Eq.~(\ref{b4b}) by $\tilde b_{2l}^*(E')$
with subsequent
subtracting the complex conjugated Eq.~(\ref{b4c}), multiplied by
$\tilde b_{1l}$. Then by integrating over $E_l$ we obtain
\begin{equation}
\dot\sigma_{bc}^{(0)}  =  i(E_2-E_1)\sigma_{bc}^{(0)}+
i\Omega_0(\sigma_{bb}^{(0)}-\sigma_{cc}^{(0)})
-\frac{1}{2}(\Gamma_L+\Gamma_R)\sigma_{bc}^{(0)}\;.
\label{bb8}
\end{equation}

Eventually we arrive to the following set of equations for $\sigma^{(n)}$
\begin{subequations}
\label{b8}
\begin{align}
\dot\sigma_{aa}^{(n)} & =  -\Gamma_L\sigma_{aa}^{(n)}
+\Gamma_R\sigma_{cc}^{(n-1)}\;,
\label{b8a}\\
\dot\sigma_{bb}^{(n)} & = \Gamma_L\sigma_{aa}^{(n)}
+\Gamma_R\sigma_{dd}^{(n-1)}+i\Omega_0(\sigma_{bc}^{(n)}-\sigma_{cb}^{(n)})\;,
\label{b8b}\\
\dot\sigma_{cc}^{(n)} & =  -\Gamma_R\sigma_{cc}^{(n)}
-\Gamma_L\sigma_{cc}^{(n)}-i\Omega_0(\sigma_{bc}^{(n)}-\sigma_{cb}^{(n)})\;,
\label{b8c}\\
\dot\sigma_{dd}^{(n)} & =  -\Gamma_R\sigma_{dd}^{(n)}
+\Gamma_L\sigma_{cc}^{(n)}\;,
\label{b8d}\\
\dot\sigma_{bc}^{(n)} & =  i(E_2-E_1)\sigma_{bc}^{(n)}+
i\Omega_0(\sigma_{bb}^{(n)}-\sigma_{cc}^{(n)})\nonumber\\
&~~~~~~~~~~~~~~~~~~~~~~~~~~~~~~~~
-\frac{1}{2}(\Gamma_L+\Gamma_R)\sigma_{bc}^{(n)}\;.
\label{b8e}
\end{align}
\end{subequations}
Using Eqs.~(\ref{b8}) we can find the charge accumulated in the collector,
$N_R(t)$, and subsequently, the total current, $I(t)=\dot N(t)$, as given by
\begin{align}
I(t)& = \sum_n n\left [\dot\sigma_{aa}^{(n)}(t)
+\dot\sigma_{bb}^{(n)}(t)
+\dot\sigma_{cc}^{(n)}(t)+\dot\sigma_{dd}^{(n)}(t)\right ]\nonumber\\
&
=\Gamma_R \left [\sigma_{cc}(t) + \sigma_{dd}(t)\right ]
\label{b9}
\end{align}
As in the previous examples, the current is
proportional to the total probability
of finding an electron in the well adjacent to the
right reservoir. The off-diagonal elements of the density matrix do
not appear in Eq.~(\ref{b9}).

Summing up over $n$ in Eqs. (\ref{b8}), we obtain the system of differential
equations for the density matrix elements of the device
\begin{subequations}
\label{b10}
\begin{align}
&\dot\sigma_{aa}  =  -\Gamma_L\sigma_{aa}
+\Gamma_R\sigma_{cc}\;,
\label{b10a}\\
&\dot\sigma_{bb}  =  \Gamma_L\sigma_{aa}
+\Gamma_R\sigma_{dd}+i\Omega_0(\sigma_{bc}-\sigma_{cb})\;,
\label{b10b}\\
&\dot\sigma_{cc}  =  -\Gamma_R\sigma_{cc}
-\Gamma_L\sigma_{cc}-i\Omega_0(\sigma_{bc}-\sigma_{cb})\;,
\label{b10c}\\
&\dot\sigma_{dd}  =  -\Gamma_R\sigma_{dd}
+\Gamma_L\sigma_{cc}\;,
\label{b10d}\\
&\dot\sigma_{bc}  =  i(E_2-E_1)\sigma_{bc}+
i\Omega_0(\sigma_{bb}-\sigma_{cc})\nonumber\\
&~~~~~~~~~~~~~~~~~~~~~~~~~~~~~~~~~~~~
-\frac{1}{2}(\Gamma_L+\Gamma_R)\sigma_{bc~}
\label{b10e}
\end{align}
\end{subequations}
Eqs.~(\ref{b10})) resemble the optical Bloch equations\cite{bloch}.
Note that the coupling with the reservoirs produces purely
negative contribution into the {\em non-diagonal}
matrix element's dynamic equation,
Eq.~(\ref{b10e}), thus causing damping of this matrix element.

Eqs.~(\ref{b10}) are solved most easily for the
stationary current, $I=I(t\to\infty )$. Using
$\sigma_{aa}+\sigma_{bb}+\sigma_{cc}+\sigma_{dd}=1$, we obtain
\begin{equation}
I=\left (\frac{\Gamma_L\Gamma_R}{\Gamma_L+\Gamma_R}\right )
\frac{\Omega_0^2}{\Omega_0^2+\Gamma_L\Gamma_R/4+\epsilon^2\Gamma_L\Gamma_R/
(\Gamma_L+\Gamma_R)^2}\;,
\label{b11}
\end{equation}
where $\epsilon =E_2-E_1$.

\subsection{Coulomb blockade.}

The extension of the rate equations (\ref{b10}) for the case of spin and
Coulomb interaction is done exactly in the same way as in Sec.~III.
Here also the rate equations for the device density matrix are
obtained only for $E_{1,2}+U$ being inside or outside the bias, but not
close to the bias edges ($\mu_R^{}\ll E_{1,2}+U\ll \mu_L^{}$ or
$E_{1,2}+U\gg \mu_L$).
Eventually we arrive to the rate equations of type Eqs.~(\ref{b10}),
but with the number of the available states of the device changed
due to additional (spin) degrees of freedom and Coulomb blockade
restrictions. The Coulomb repulsion
manifests itself also in a modification of the
transition amplitude $\Omega$ and the rates $\Gamma$'s, Eq.~(\ref{cc3}).

In the case of large Coulomb repulsion, some of electron states
of the device are outside the bias (the Coulomb blockade). As a result,
the number of the equations is reduced.
Consider, for instance, the situation where the Coulomb interaction $U$ of two
electrons in the same well so large that $E_{1,2}+U\gg\mu_L^{}$, but
the Coulomb repulsion of two electrons in different wells, $\bar U$, is
much smaller, so that $E_{1,2}+\bar U\ll \mu_L^{}$.
Then the state of two electrons in the same well is not available, but
two electrons can occupy different wells. In this case the rate
equations for the corresponding density matrix elements of the device are
\begin{subequations}
\label{b12}
\begin{align}
& \dot\sigma_{aa} = -2\Gamma_L\sigma_{aa}
+\Gamma_R(\sigma_{cc\uparrow}+\sigma_{cc\downarrow})\;,
\label{b12a}\\
& \dot\sigma_{bb\uparrow} = \Gamma_L\sigma_{aa}
+\Gamma'_R(\sigma_{dd\uparrow\uparrow}+\sigma_{dd\uparrow\downarrow})\nonumber\\
&~~~~~~~~~~~~~~~~~~~~~~~~~~~~~~~~~~
+i\Omega_0(\sigma_{bc\uparrow}-\sigma_{cb\uparrow})\;,
\label{b12b}\\
& \dot\sigma_{cc\uparrow} = -\Gamma_R\sigma_{cc\uparrow}
-2\Gamma'_L\sigma_{cc\uparrow}
-i\Omega_0(\sigma_{bc\uparrow}-\sigma_{cb\uparrow})\;,
\label{b12c}\\
& \dot\sigma_{dd\uparrow\uparrow} = -\Gamma'_R\sigma_{dd\uparrow\uparrow}
+\Gamma'_L\sigma_{cc\uparrow}\;,
\label{b12d}\\
& \dot\sigma_{bc\uparrow} = i(E_2-E_1)\sigma_{bc\uparrow}+
i\Omega_0(\sigma_{bb\uparrow}-\sigma_{cc\uparrow})\nonumber\\
&~~~~~~~~~~~~~~~~~~~~~~~~~~~~~~~~~~
-\frac{1}{2}(2\Gamma'_L+\Gamma_R)\sigma_{bc\uparrow}\;,
\label{b12e}
\end{align}
\end{subequations}
where $\Gamma'_{L(R)}=2\pi\rho_{L(R)}(E_1+\bar U)
|\Omega_{L(R)}(E_1+\bar U)|^2$. Here for the shortness
we wrote only the equations for the ``spin up'' component of
the density matrix. The same equations are obtained for the
``spin down'' components of the density matrix. The total current is
\begin{equation}
I=\Gamma_R(\sigma_{cc\uparrow}+\sigma_{cc\downarrow})+
\Gamma'_R(\sigma_{dd\uparrow\uparrow}+\sigma_{dd\uparrow\downarrow}+
\sigma_{dd\downarrow\uparrow}+\sigma_{dd\downarrow\downarrow}).
\label{b13}
\end{equation}

It is quite clear that the ``spin up'' and ``spin down''
components of the density matrix are equal, i.e.
$\sigma_{bb\uparrow} = \sigma_{bb\downarrow}
= \sigma_{bb}$, the same holding for $\sigma_{cc}$, $\sigma_{dd}$
components. Therefore
Eqs. (\ref{b12}), (\ref{b13}) can be rewritten as
\begin{subequations}
\label{b14}
\begin{align}
\dot\sigma_{aa} & =  -2\Gamma_L\sigma_{aa}
+2\Gamma_R\sigma_{cc}\;,
\label{b14a}\\
\dot\sigma_{bb} & =  \Gamma_L\sigma_{aa}
+2\Gamma'_R\sigma_{dd}+i\Omega_0(\sigma_{bc}-\sigma_{cb})\;,
\label{b14b}\\
\dot\sigma_{cc} & =  -\Gamma_R\sigma_{cc}
-2\Gamma'_L\sigma_{cc}-i\Omega_0(\sigma_{bc}-\sigma_{cb})\;,
\label{b14c}\\
\dot\sigma_{dd} & =  -\Gamma'_R\sigma_{dd}
+\Gamma'_L\sigma_{cc}\;,
\label{b14d}\\
\dot\sigma_{bc} & =  i(E_2-E_1)\sigma_{bc}+
i\Omega_0(\sigma_{bb}-\sigma_{cc})~\nonumber\\
&~~~~~~~~~~~~~~~~~~~~~~~~~~~~~~~~
-\frac{1}{2}(2\Gamma'_L+\Gamma_R)\sigma_{bc}\;,
\label{b14e}
\end{align}
\end{subequations}
and
\begin{equation}
I=2\Gamma_R\sigma_{cc}+4\Gamma'_R\sigma_{dd}
\label{b15}
\end{equation}

Using $\sigma_{aa}+2\sigma_{bb}+2\sigma_{cc}+4\sigma_{dd}=1$ we obtain for
the dc current
\begin{align}
&I=\left (\frac{2\Gamma_L\Gamma'_R}{2\Gamma'_L+\Gamma_R}\right )\nonumber\\
&
\frac{\Omega_0^2}{4\Omega_0^2\frac{\Gamma_L\Gamma'_L+
\Gamma_L\Gamma'_R+\Gamma_R\Gamma'_R/4}
{(2\Gamma'_L+\Gamma_R)^2}
+\frac{\Gamma_L\Gamma'_R}{2}+\epsilon^2\frac{2\Gamma_L\Gamma'_R}
{(2\Gamma'_L+\Gamma_R)^2}}\;,
\label{b16}
\end{align}
where $\epsilon =E_2-E_1$. Notice that the
current (\ref{b16}) differs from that given by Eq.~(\ref{b11})
even for $\Gamma'_L=\Gamma_L$ and  $\Gamma'_R=\Gamma_R$,
despite the fact that in the both cases only one electron can occupy
each of the wells.

It is interesting to compare our result with that of Stoof and
Nazarov\cite{naz1} for the case of strong Coulomb repulsion
between two electrons in different wells ($E_{1,2}+\bar U\gg E_F^L$),
where only one
electron can be found inside the system. It corresponds
to $\Gamma'_L=0$. In this case the dc current given by  Eq.~(\ref{b16}) is
\begin{equation}
I=\frac{\Gamma_R\Omega_0^2}{\displaystyle\Omega_0^2(2+\Gamma_R/2\Gamma_L)+
\Gamma_R^2/4+\epsilon^2}.
\label{b17}
\end{equation}
This result is slightly different from that obtained by Stoof and
Nazarov (by the factor two in front of $\Gamma_L$). The difference stems from
the account of spin components in the rate equations, which has
not been done in\cite{naz1}.

\section{Inelastic processes}

\begin{figure}[tbh]
\includegraphics[width=7cm]{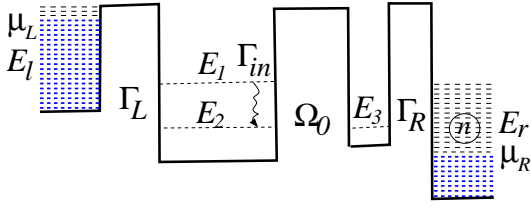}
\caption{Resonant tunneling through a double-well structure in the presence of inelastic process.}
\label{fig21}
\end{figure}

As an example of a system with coherent tunneling
accompanied by inelastic scattering, let us consider
the coupled-dot structure shown in Fig.~\ref{fig21}.
In this system a resonant current flows
due to inelastic transition from the upper to the lower level
in the left well. For simplicity, we restrict ourselves to non-interacting
spin-less electrons. The Coulomb interaction and the spin effects
can be accounted for precisely in the same way as we did in the previous
sections.

The tunneling Hamiltonian of the system has the following structure
\begin{align}
H& =  \sum_l E_{l}a^{\dagger}_{l}a_{l} +
              E_1 a_1^{\dagger}a_{1} + E_2 a_2^{\dagger}a_{2}+
              E_3 a_3^{\dagger}a_{3}\nonumber\\
              &+
              \sum_{\alpha} E_{\alpha}^{ph}c^{\dagger}_{\alpha}c_{\alpha}+
              \sum_r E_{r}a^{\dagger}_{r}a_{r}+
              \Omega_0(a_2^{\dagger}a_{3}+a_3^{\dagger}a_{2})
\nonumber\\
           &+ \sum_l \Omega_{l}(a^{\dagger}_{l}a_1 +
              a^{\dagger}_{1}a_{l})+
   \sum_\alpha \Omega_{\alpha}^{ph}(a^{\dagger}_{2}a_{1}c_{\alpha}^{\dagger} +
              a^{\dagger}_{1}a_{2}c_{\alpha})\nonumber\\
              &+
              \sum_r \Omega_{r}(a^{\dagger}_{r}a_3+
              a^{\dagger}_{3}a_{r}).
\label{e1}
\end{align}
Here the subscript $\alpha$ enumerates the states in the phonon bath
and $\Omega_{\alpha}^{ph}$ is the corresponding coupling. The
many particle time-dependent wave function of the system is
\begin{align}
&|\Psi (t)\rangle  =  \left [ b_0(t) + \sum_l b_{1l}(t)
     a_{1}^{\dagger}a_{l}+ \sum_{l,\alpha} b_{2l\alpha}(t)a_{2}^{\dagger}
a_{l}c_{\alpha}^{\dagger}
\right.\nonumber\\
      &\left.
~~~~~~+\sum_{l,\alpha} b_{3l\alpha}(t)a_{3}^{\dagger}
a_{l}c_{\alpha}^{\dagger}+ \sum_{l<l',\alpha} b_{12ll'\alpha}(t)
           a_{1}^{\dagger}a_{2}^{\dagger}a_{l}a_{l'}c_{\alpha}^{\dagger}
           \right.
      \nonumber\\
      &\left.~~~~~~ +\sum_{l<l',\alpha} b_{13ll'\alpha}(t)
           a_{1}^{\dagger}a_{3}^{\dagger}a_{l}a_{l'}c_{\alpha}^{\dagger}
           + \ldots \right ] |0\rangle.
\label{e2}
\end{align}
Repeating the procedure of the previous sections we find the following set
of equations for the Laplace transformed amplitudes, $\tilde b(E)$:
\begin{subequations}
\label{e3}
\begin{align}
& (E + i\Gamma_L/2) \tilde{b}_{0}=i
\label{e3a}\\
& (E + E_{l} - E_1+i\Gamma_{in}/2) \tilde{b}_{1l}-\Omega_{l} \tilde b_0=0
\label{e3b}\\
& (E + E_{l} - E_{\alpha}-E_2+i\Gamma_L/2) \tilde{b}_{2l\alpha} -
      \Omega_{\alpha}^{ph}\tilde b_{1l}\nonumber\\
      &~~~~~~~~~~~~~~~~~~~~~~~~~~~~~~~~~~~~~~~~~~~~~
-\Omega_{0} \tilde{b}_{3l\alpha}=0
\label{e3c}\\
& (E + E_{l} -E_{\alpha}-E_3+i\Gamma_L/2+i\Gamma_R/2) \tilde{b}_{3l\alpha} \nonumber\\
      &~~~~~~~~~~~~~~~~~~~~~~~~~~~~~~~~~~~~~~~~~~~~~
-\Omega_{0} \tilde{b}_{2l\alpha}=0
\label{e3d}\\
& (E + E_{l} + E_{l'} - E_1 - E_2 -E_{\alpha})\tilde{b}_{12ll'\alpha} -
       \Omega_{l'} \tilde{b}_{2l\alpha}
       \nonumber\\
      &~~~~~~~~~~~~~~~~~~~~~~~~~~~~~~
       +\Omega_{l} \tilde{b}_{2l'\alpha}
       -\Omega_{0}\tilde{b}_{13ll'\alpha}=0
\label{e3e}\\
&\big(E + E_{l} + E_{l'} - E_1 - E_3 -E_{\alpha}+ i\Gamma_{in}/2+ i\Gamma_R/2\big)\tilde{b}_{13ll'\alpha}
 \nonumber\\
&~~~~~~~~~~~~~~~~
-\Omega_{0}\tilde{b}_{12ll'\alpha}-\Omega_{l'}\tilde{b}_{3l\alpha}
+\Omega_{l}\tilde{b}_{3l'\alpha}=0
\label{e3f}\\
&~~~~~~~~~~~~~~~~~~~~~~~~~~\cdots\nonumber
\end{align}
\end{subequations}
where $\Gamma_{in}=2\pi\rho_{ph}|\Omega^{ph}|^2$ is the partial width
of the level $E_1$ due to phonon emission and $\rho_{ph}$ is the density of
phonon states.

\begin{figure}[tbh]
\includegraphics[width=7cm]{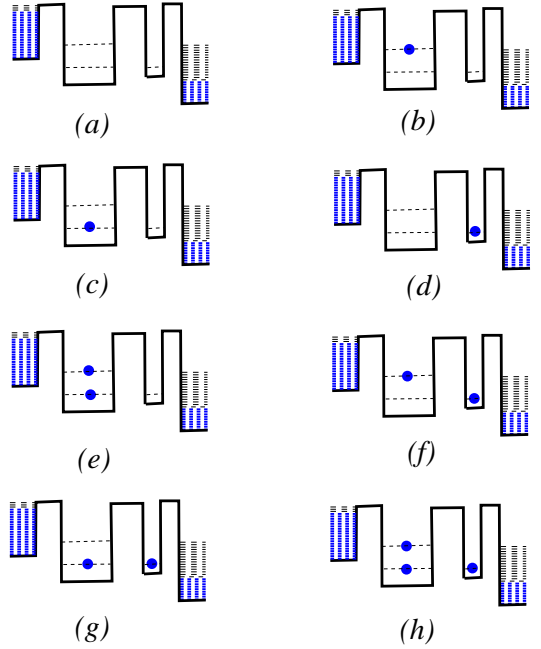}
\caption{All possible electron states of the device, shown in Fig.~\ref{fig21}: (a) — all the levels $E_{1,2,3}$ are empty; (b) — the upper level, $E_1$, is occupied; (c) — the lower level, $E_2$, is occupied; (d) — the level $E_3$ is occupied; (e) — the levels $E_1$ and $E_2$ are occupied; (f) — the levels $E_1$ and $E_3$ are occupied; (g) — the levels $E_2$ and $E_3$ are
occupied; (h) — all the levels $E_{1,2,3}$ are occupied.}
\label{fig22}
\end{figure}

The density matrix elements of the device is $\sigma_{ij}(t)=
\sum_n\sigma_{ij}^{(n)}(t)$, where $\sigma_{ij}^{(n)}(t)$, are
related to the amplitudes $\tilde b(E)$ via Eqs.~(\ref{invlap}), (\ref{invlapp})
All possible electron states of the device are shown
in Fig.~\ref{fig22}.
Using the previous section procedure for {\em diagonal}
matrix elements we obtain master equations analogous to Eq.~(\ref{b10}),
in which transitions between isolated levels $E_2$ and $E_3$ take place
through the coupling with non-diagonal matrix elements. These equations have the appearance of the optical Bloch equation\cite{bloch}. However, the master equation for the {\em non-diagonal} matrix element (coherences)
contains an additional term. Therefore, we present the derivation of
the master equations for ``coherences'' $\sigma_{ef}$ and
$\sigma_{cd}$ in some detail.

Consider for example
the non-diagonal density submatrix elements
$\sigma_{cd}^{(0)}=\sum_{l,\alpha}b_{2l\alpha}(t)b^*_{3l\alpha}(t)$ and
$\sigma_{ef}^{(0)}=\sum_{l<l',\alpha}b_{12ll'\alpha}(t)b^*_{13ll'\alpha}(t)$.
The differential equation for $\sigma_{cd}^{(0)}(t)$
can be obtained by multiplying Eq.~(\ref{e3c}) by $\tilde b^*_{3l\alpha}(E')$
with subsequent subtraction of the complex conjugated Eq.~(\ref{e3d}),
multiplied by $\tilde b_{2l\alpha}(E)$. Then using Eq.~(\ref{invlap}), (\ref{invlapp})
we obtain
\begin{equation}
\dot\sigma_{cd}^{(0)}  =  i(E_3-E_2)\sigma_{cd}^{(0)}+
i\Omega_0(\sigma_{cc}^{(0)}-\sigma_{dd}^{(0)})
-\frac{1}{2}(2\Gamma_L+\Gamma_R)\sigma_{cd}^{(0)}\;.
\label{e4}
\end{equation}
Similarly, multiplying Eq.~(\ref{e3e}) by
$\tilde b^*_{13ll'\alpha}(E')$ and  Eq.~(\ref{e3f})
by $\tilde b_{12ll'\alpha}(E)$, we find
the differential equation for $\sigma_{ef}^{(0)}(t)$
\begin{align}
\dot\sigma_{ef}^{(0)} &=  i(E_3-E_2)\sigma_{ef}^{(0)}+
i\Omega_0(\sigma_{ee}^{(0)}-\sigma_{ff}^{(0)})\nonumber\\
&~~~~~~~~~~~~~~~~~~~~~~~~~~
-\frac{1}{2}(\Gamma_{in}+\Gamma_R)\sigma_{ef}^{(0)}-iF
\label{e5}
\end{align}
where
\begin{align}
F&=\sum_{l<l',\alpha}\int\frac{dEdE'}{4\pi^2}\left [
\tilde{b}^*_{13ll'\alpha}(E')\Omega_{l'}\tilde b_{2l\alpha}(E)\right.\nonumber\\
&\left.
-\tilde{b}^*_{13ll'\alpha}(E')\Omega_{l}\tilde b_{2l'\alpha}(E)-\tilde{b}_{12ll'\alpha}(E)\Omega_{l'}\tilde b^*_{3l\alpha}(E')\right.
\nonumber\\
&\left.
+\tilde{b}_{12ll'\alpha}(E)\Omega_{l}\tilde b^*_{3l'\alpha}(E')\right ]
e^{i(E'-E)t}
\label{e6}
\end{align}
Substituting the amplitudes $\tilde{b}_{12ll'\alpha}$
from Eq.~(\ref{e3e}) and
$\tilde{b}^*_{13ll'\alpha}$ from Eq.~(\ref{e3f}) into Eq.~(\ref{e6}),
and replacing the sum over $l(l')$ by the corresponding integral,
we find $-iF=\Gamma_L\sigma_{cd}^{(0)}$. It implies that
the non-diagonal density matrix $\sigma_{ef}$ given by Eq.~(\ref{e5}),
is coupled with $\sigma_{cd}$ via a single electron transition
from the emitter to the left well. Obviously, such a term does not appear in the Bloch
equations, which deal with two-level systems.

Summing up over $n$ in the rate equations for the density submatrix
$\sigma^{(n)}_{ij}(t)$ we obtain the set of rate equations for
the density matrix of the device
\begin{subequations}
\label{e8}
\begin{align}
\dot\sigma_{aa} & =  -\Gamma_L\sigma_{aa}+\Gamma_{R}\sigma_{dd}
\label{e8a}\\
\dot\sigma_{bb} & =  \Gamma_{L}\sigma_{aa}-\Gamma_{in}
\sigma_{bb}+\Gamma_{R}\sigma_{ff}
\label{e8b}\\
\dot\sigma_{cc} & =  \Gamma_{in}\sigma_{bb}+i\Omega (\sigma_{cd}-\sigma_{dc})
+\Gamma_R\sigma_{gg}-\Gamma_{L}\sigma_{cc}
\label{e8c}\\
\dot\sigma_{dd} & =  -\Gamma_R\sigma_{dd}+i\Omega (\sigma_{dc}-\sigma_{cd})
-\Gamma_{L}\sigma_{dd}
\label{e8d}\\
\dot\sigma_{ee} & =  \Gamma_L\sigma_{cc}+i\Omega (\sigma_{ef}-\sigma_{fe})
+\Gamma_{R}\sigma_{hh}
\label{e8e}\\
\dot\sigma_{ff} & =  \Gamma_L\sigma_{dd}-\Gamma_{R}\sigma_{ff}
+i\Omega (\sigma_{fe}-\sigma_{ef})-\Gamma_{in}\sigma_{ff}
\label{e8f}\\
\dot\sigma_{gg} & =  \Gamma_{in}\sigma_{ff}-\Gamma_{R}\sigma_{gg}
-\Gamma_L\sigma_{gg}
\label{e8g}\\
\dot\sigma_{hh} & =  \Gamma_L\sigma_{gg}-\Gamma_{R}\sigma_{hh}\;,
\label{e8h}\\
\dot\sigma_{cd} & =  i(E_3-E_2)\sigma_{cd}+
i\Omega (\sigma_{cc}-\sigma_{dd})\nonumber\\
&~~~~~~~~~~~~~~~~~~~~~~~~~~
-1/2(2\Gamma_L+\Gamma_R)
\sigma_{cd}
\label{e8i}\\
\dot\sigma_{ef} & =  i(E_3-E_2)\sigma_{ef}+
i\Omega (\sigma_{ee}-\sigma_{ff})\nonumber\\
&~~~~~~~~~~~~~~~~~~~~
-1/2(\Gamma_{in}+
\Gamma_R)\sigma_{ef}+\Gamma_L\sigma_{cd}
\label{e8j}
\end{align}
\end{subequations}
and the resonant current flowing through this system is
$I=\Gamma_R[\sigma_{dd}+\sigma_{ff}+\sigma_{gg}+\sigma_{hh}]$.

\section{Master equations for two wells separated by continuum}

\begin{figure}[tbh]
\includegraphics[width=8cm]{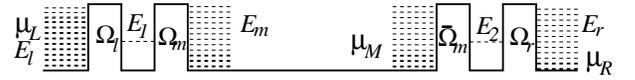}
\caption{(Color online) Resonant transport through two quantum dots separated
by a ballistic channel. Here $\Omega_l$ and $\Omega_r$ denote the coupling of
the left and right dots dot with the levels $E_l$ and $E_r$ in the
left and in the right leads. $\Omega_m$ and $\bar\Omega_m$ denote
the coupling of the left and the right dots with the level $E_m$ in
the ballistic channel.}
\label{fig3}
\end{figure}

Let us consider quantum transport through two quantum wells (quantum dots). separated by a ballistic channel, as shown schematically in Fig.~\ref{fig3}.
The dots contain only isolated levels, whereas
the density of states in the ballistic channel and in the emitter and the
detector is very high (continuum).
This system can be described by the tunneling Hamiltonian
\begin{align}
H&=\sum_l E_{l}a^{\dagger}_{l}a_{l} +
E_1 a_1^{\dagger}a_{1}+
\sum_m E_{m}a^{\dagger}_{m}a_{m} +
E_2 a_2^{\dagger}a_{2}\nonumber\\
&+\sum_r a_{r}a^{\dagger}_{r}a_{r}+\sum_{i,j=1,2}U_{ij}n_in_j+ \left\{ \sum_l \Omega_{l}a^{\dagger}_{1}a_l\right.\nonumber\\
&\left. + \sum_m \Omega_{m}a^{\dagger}_ma_{1}
+ \sum_m \bar\Omega_{m}a^{\dagger}_{2}a_m
+ \sum_r \Omega_{r}a_r^{\dagger}a_2 + {\rm H.c.}\right\}
\label{a1}
\end{align}
where $n_i=a_i^\dagger a_i^{}$. The subscripts $l$, $m$ and
$r$ enumerate correspondingly the levels in the
left reservoir, in the (middle)
ballistic channel and in the right reservoir.
The spin degrees of freedom were omitted.

In order to simplify the derivation we assumed that the intradot charging
energy $U_{ii}$ is large, $E_{1,2}+U_{ii}\gg \mu_L$. Thus
only one electron can occupy each of the dots.
However, the interdot charging energy $U_{12}$
is much smaller, so it does not prevent simultaneous occupation
of the two dots. The same is assumed for
the Coulomb repulsion between electrons inside the dots and
the ballistic channel. Although we did not include this interaction
in the Hamiltonian (\ref{a1}), it can be treated in the same way
as the interdot interaction $U_{12}$.
As in the previous case we restrict ourselves to the zero temperature case,
even though the results are valid for a finite temperature, as
would be clear from the derivation.

\begin{figure}[tbh]
\includegraphics[width=8cm]{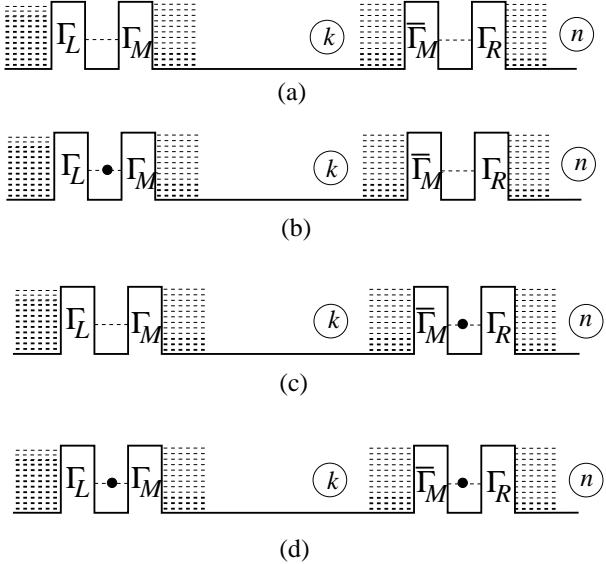}
\caption{Electron states of the two separated dot structure, shown in Fig.~\ref{fig3}. $\Gamma_{L,R}$, $\Gamma_M$ and $\bar\Gamma_M$ are the tunneling rates between the dots and the reservoirs, and between the dots and the
ballistic channel. The indices $k$ and $n$ denote the number of electrons
penetrating to the ballistic channel and to the collector at time $t$.}
\label{fig4}
\end{figure}

Let us assume that all the levels in the emitter, in the
ballistic channel and in the collector are initially filled up to the Fermi
energies $\mu_L$, $\mu_M$ and $\mu_R$ respectively. We call
it as the ``vacuum'' state, $|0\rangle$. (In the following we consider
the case of large bias, so that $\mu_L\gg \mu_M\gg \mu_R$).
The many-body wave function describing this system can be written
in the occupation number representation as
\begin{align}
&\Psi (t)\rangle  =  \left [ b_0(t) + \sum_l b_{1l}(t)
     a_{1}^{\dagger}a_{l} + \sum_{l,m} b_{lm}(t)a_{m}^{\dagger}a_{l}\right.\nonumber\\
    &\left.+\sum_l b_{2l}(t)a_{2}^{\dagger}a_{l}
      + \sum_{l,r} b_{lr}(t)a_{r}^{\dagger}a_{l}
           +\sum_{l<l'} b_{12ll'}(t)a_{1}^{\dagger}a_{2}^{\dagger}a_{l}a_{l'}\right.
           \nonumber\\
           &\left.~~~~~~~~~~~~~~~~~~+\sum_{l<l',r} b_{1ll'r}(t)
           a_{1}^{\dagger}a_{r}^{\dagger}a_{l}a_{l'}
           + \ldots \right ] |0\rangle,
\label{a2}
\end{align}
where $b(t)$ are the time-dependent probability amplitudes to
find the system in the corresponding states described above.
These amplitudes are obtained from the Shr\"odinger equation
$i|\dot\Psi (t)\rangle =H|\Psi (t)\rangle$, supplemented
with the initial condition
($b_0(0)=1$, and all the other $b(0)$'s being zeros). Using the amplitudes
$b(t)$ we can find the density-matrix of the quantum dots,
$\sigma_{ij}^{(k,n)}(t)$, by tracing out the continuum
states of the reservoirs and the ballistic channel. Here the subscript
indices in $\sigma$
denote four states of the dots:  $i,j=\{a,b,c,d\}$, where
$|a\rangle$ -- the levels $E_{1,2}$ are empty,
$|b\rangle$ -- the level $E_1$ is occupied,
$|c\rangle$ -- the level $E_2$ is occupied,
$|d\rangle$ -- the both level $E_{1,2}$ are occupied,
and the superscript indices $k,n$ denote the number of
electrons accumulated in the ballistic channel and in the collector
respectively at time $t$, Fig.~\ref{fig4},

One finds
\begin{subequations}
\label{a7}
\begin{align}
&\sigma_{aa}^{(0,0)}(t) = |b_{0}(t)|^2,~~~
\sigma_{aa}^{(1,0)}(t)= \sum_{l,m} |b_{lm}(t)|^2,\nonumber\\
&
\sigma_{aa}^{(1,1)}(t)=\sum_{l<l',m,r} |b_{ll'mr}(t)|^2,~~~ \ldots
\label{a7a}\\
&\sigma_{bb}^{(0,0)}(t) =\sum_l |b_{1l}(t)|^2,~~~
\sigma_{bb}^{(1,0)}(t)=\sum_{l<l',m} |b_{1ll'm}(t)|^2,
\nonumber\\
&\sigma_{bb}^{(1,1)}(t)=\sum_{l<l'<l'',m,r} |b_{1ll'l''mr}(t)|^2,~~~ \ldots
\label{a7b}\\
&\sigma_{bc}^{(0,0)}(t) = \sum_l b_{1l}(t)b^*_{2l}(t),\nonumber\\
&
\sigma_{bc}^{(1,0)}(t)= \sum_{l<l',m} b_{1ll'm}(t)b^*_{2ll'm}(t),
\nonumber\\
&\sigma_{bc}^{(1,1)}(t)=\sum_{l<l'<l'',m,r} b_{1ll'l''mr}(t)b^*_{2ll'l''mr}(t),~~\ldots
\label{a7c}\\
&~~~~~~~~~~~~~~~~~~~~~~~~~~~~~~~~~~~~~~\cdots\nonumber
\end{align}
\end{subequations}

The rate of electrons arriving to the collector determines the electron
current in the system. Therefore the current operator is
$\hat I=i[H,\hat N_R]$, where
$\hat N_R=\sum_ra^{\dagger}_ra_r$ is the operator for
the total number of electrons accumulated in the right reservoir.
Using Eqs.~(\ref{a1}), (\ref{a2}), (\ref{a7}) we find that the
current $I(t)$ flowing through the system is
\begin{align}
&I(t)=\langle\Psi (t)|\hat I|\Psi (t)\rangle\nonumber\\
&=
\sum_{k,n} n\left [\dot\sigma_{aa}^{(k,n)}(t)+
\dot\sigma_{bb}^{(k,n)}(t)+
\dot\sigma_{cc}^{(k,n)}(t)+
\dot\sigma_{dd}^{(k,n)}(t)\right ]
\label{a8}
\end{align}
As expected, $I(t)$ is the time derivative of the total charge accumulated in the collector. Thus the current $I(t)$ flowing through this system is expressed in terms of the diagonal elements of the density-matrix $\sigma (t)$. In order to find the differential equations for $\sigma (t)$ we need to sum over the states of the reservoirs and the ballistic channel, Eqs.~(\ref{a7}).

As in previous sections, we use the Laplace transform for the Shr\"odinger equation, $i|\dot\Psi (t)\rangle =H|\Psi (t)\rangle$. Then the amplitudes $b(t)$ in the wave function (\ref{a2}) are replaced
by their Laplace transform, $\tilde{b}(E)$, Eq.~(\ref{lap}).
Substituting Eq.~(\ref{a2}) into the Shr\"odinger equation
we obtain an infinite set of coupled equations for the
amplitudes $\tilde b(E)$:
\begin{subequations}
\label{aa4}
\begin{align}
&E \tilde{b}_{0}(E) - \sum_l \Omega_{l}\tilde{b}_{1l}(E)=i
\label{aa4a}\\
&(E + E_{l} - E_1) \tilde{b}_{1l}(E) - \Omega_{l}
      \tilde{b}_0(E) -\sum_m\Omega_{m}\tilde{b}_{lm}(E)=0
\label{aa4b}\\
&(E + E_{l} - E_m) \tilde{b}_{lm}(E)-\Omega_m\tilde{b}_{1l}(E)
-\bar\Omega_m\tilde{b}_{2l}(E)\nonumber\\
&~~~~~~~~~~~~~~~~~~~~~~~~~~~~~~~~~-\sum_{l'}\Omega_{l'}\tilde{b}_{12ll'}(E)=0
\label{aa4c}\\
&(E + E_{l} - E_2) \tilde{b}_{2l}(E) -
      \sum_m\bar\Omega_m \tilde{b}_{lm}(E) \nonumber\\
&~~~~~~~~~~~~
-\sum_{r} \Omega_{r}\tilde{b}_{lr}(E)-
      \sum_{l'} \Omega_{l'}\tilde{b}_{12ll'}(E)=0
\label{aa4d}\\
&(E + E_{l} + E_{l'} - E_1 - E_2-U_{12}) \tilde{b}_{12ll'}(E)-
\Omega_{l'} \tilde{b}_{2l}(E)
\nonumber\\
&~~~+
\Omega_{l} \tilde{b}_{2l'}(E)
-\sum_{m} \bar\Omega_{m}\tilde{b}_{1ll'm}(E)\nonumber\\
&~~~~~~~~
-\sum_{m} \Omega_{m}\tilde{b}_{2ll'm}(E)
-\sum_{r} \Omega_{r}\tilde{b}_{1ll'r}(E)=0
\label{aa4e}\\
&~~~~~~~~~~~~~~~~~~~~~~~~~~~~~~~~~~~\cdots
\nonumber
\end{align}
\end{subequations}
Note that due to the Pauli principle an electron can return back
only into unoccupied states of the emitter. As a result, the summation
over the emitter states does not appear in the corresponding terms
of Eqs.~(\ref{aa4}) (in the second term of Eq.~(\ref{aa4b}),
in the second and the third terms of Eq.~(\ref{aa4c}), and so on).

Now we replace the amplitude $\tilde b$ in the term $\sum\Omega\tilde b$
of each of the equations~(\ref{aa4}) by its expression obtained from the subsequent equation. For example, substitute $\tilde{b}_{1l}(E)$ from Eq.~(\ref{aa4b}) into Eq.~(\ref{aa4a}). We obtain
\begin{align}
&\left [ E - \sum_l \frac{\Omega^2_L(E_l)}{E + E_{l} - E_1}
    \right ] \tilde{b}_{0}(E)\nonumber\\
&~~~~~~~~~~~~~~~~
- \sum_{l,m}
    \frac{\Omega_L(E_l)\Omega_M(E_m)}{E + E_{l} - E_1}
    \tilde{b}_{lm}(E)=i,
\label{a5}
\end{align}
where $\Omega_l\equiv \Omega_L(E_l)$ and $\Omega_m\equiv \Omega_M(E_m)$. In the wide-band limit, this expression is treated exactly in the same way as Eq.~(\ref{exam}). Namely,
we replace the sums over $l$ and $m$ by integrals, like
$\sum_{l}\;\rightarrow\;\int \rho_{L}(E_{l})\,dE_{l}\:$,
where $\rho_{L}(E_{l})$ is the density of states in the emitter.
Then the first sum in Eq.~(\ref{a5}) becomes an
integral which can be split into a sum of the singular and principal value
parts. The singular part
yields $\;\,-i\Theta (\mu_L+E-E_1)\,\Gamma_L/2$, where $\Gamma_L = 2\pi
\rho_L(E_1)|\Omega_L(E_1)|^2$ is the level $E_1$ partial width
due to coupling to the emitter. In the large bias limit,
$\mu_L\gg E_1\gg \mu_M$, the integration over $E_{l(m)}$-variables can be extended to $\pm\infty$. As a result, the theta-function can be replaced by one, whereas the principal p[art vanishes (see Eqs.~(\ref{exam1}), (\ref{exam2})). The second sum (integral) in Eq.~(\ref{a5}) vanishes as well, since the poles of the integrand in the $E_l$-variable are on one side of the integration contour (c.f. with Eqs.~(\ref{exam3})-(\ref{eexam5})). In general, any terms of the type $\int\cdots dE_s\cdots \tilde b(\cdots ,E_s,\cdots )
(E+\cdots\pm E_s)^{-1}\to 0$, whenever the integration over
the $E_s$-variable can be extended to $\pm \infty$. We shall imply
this property in all subsequent derivations. Notice
that these results are valid also for non-zero temperature, providing
that $T\ll \mu_L-E_{1,2},\; E_{1,2}-\mu_M$.

Now we apply analogous considerations to the other equations of the
system (\ref{aa4}). However, in order to be able to carry it, we need to impose some restriction on a geometry of the ballistic channel, separated two wells, Fig.~\ref{fig3}. This already appears by treating the last term of Eq.~(\ref{aa4b}), which we denote as $S_M=\sum_m\Omega_m\tilde b_{lm}$. Substituting the amplitude $\tilde{b}_{lm}(E)$, obtained from Eq.~(\ref{aa4c}) into this term and replacing the sum by integral, we find
\begin{align}
S_M=\tilde  b_{2l}(E)\int\limits_{-\Lambda}^\Lambda {\Omega_M^{}(E_m)\bar\Omega_M^{}(E_m)\rho_M^{}\over E+E_l-E_m}dE_m
\label{coh}
\end{align}
In the large band limit, the spectral functions $\Omega_M^2(E_m)\rho_M^{}(E_m)$ and $\bar\Omega_M^2(E_m)\rho_M^{}(E_m)$ are weakly dependent of energy $E_m$. However the relative sign of $\Omega_M^{}(E_m)$ and $\bar \Omega_M^{}(E_m)$ can strongly oscillate with $E_m$.

Indeed the tunneling couplings are given by the overlap of localized wave functions, belonging to the dot states, $E_{1,2}$, with an extended wave function, belonging to the reservoir state $E_m$. Since the latter  belongs to continuum, the corresponding wave function would oscillate inside the middle reservoir. Then the sign[$\Omega_{M}(E_m)\bar\Omega_{M}(E_m)$] will oscillate as well with a frequency $\sim E_m^{1/2}L$, where $L$ is a length of the reservoir. This creates a problem of how to make the product of the  both couplings energy independent. in order to perform the $E_m$-integration in the same way, as we did in Eqs.~(\ref{exam}), (\ref{a5}).

\begin{figure}[tbh]
\includegraphics[width=8cm]{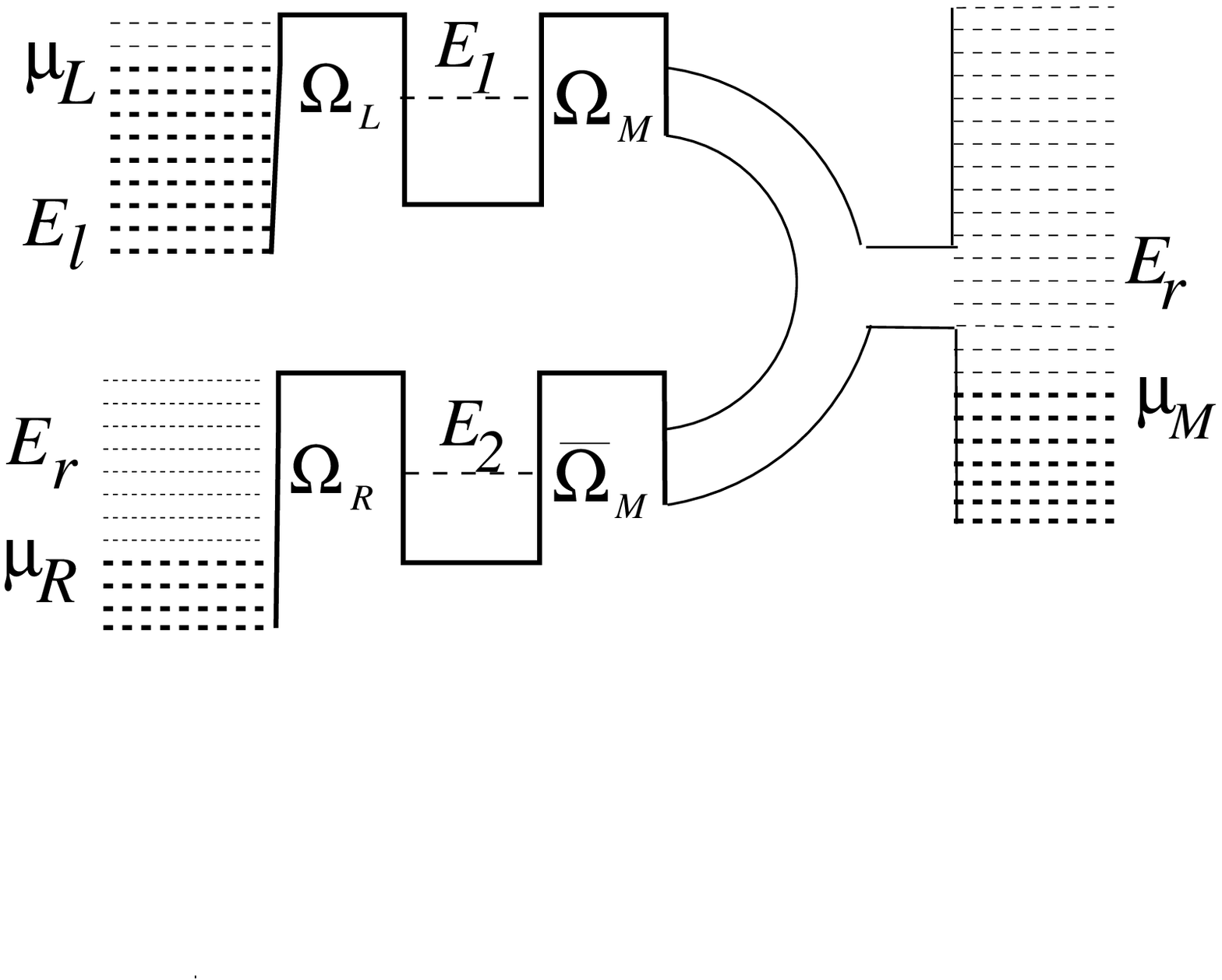}
\caption{Two quantum dots, separated by a common reservoir (ballistic channel), as in Fig.~\ref{fig3}, where the dots are coupled to the reservoir at spacially close points.}
\label{fig5}
\end{figure}

The problem can be avoided by coupling two quantum dots to a common reservoir (ballistic channel) at close points, similar to the setup in Refs.~[\onlinecite{wolf,xq1,xq2}] and shown in Fig.~\ref{fig5}.
This would make the product $\Omega_{M}(E_m) \bar\Omega_{M}(E_m)\rho_M$ independent of the energy $E_m$ in the  wide-band limit. Thus $\Omega_{M}(E_m) \bar\Omega_{M}(E_m)\to \Omega_{M} \bar\Omega_{M}$, so that the integration in Eq.~(\ref{coh}) and in similar terms can be performed as in Eqs.~(\ref{exam1}), (\ref{exam2}), finally arriving at the following set of equations:
\begin{subequations}
\label{a6}
\begin{align}
&(E + i\Gamma_L/2) \tilde{b}_{0}=i
\label{a6a}\\
&(E + E_{l} - E_1 + i\Gamma_M/2) \tilde{b}_{1l}
\nonumber\\
&~~~~~~~~~~~~~~~~~~~~~
- \Omega_{l} \tilde{b}_{0}
+i\pi\rho_M\Omega_M\bar\Omega_M\tilde{b}_{2l}=0
\label{a6b}\\
&(E + E_{l} - E_{m} + i\Gamma_L/2) \tilde{b}_{lm} -
      \Omega_{m} \tilde{b}_{1l}-\bar\Omega_{m} \tilde{b}_{2l}=0
\label{a6c}\\
&(E + E_{l} -E_2+ i\Gamma_L/2+ i\bar\Gamma_M/2+ i\Gamma_R/2)
       \tilde{b}_{2l}
\nonumber\\
&~~~~~~~~~~~~~~~~~~~~~~~~~~~~~~~~~
+i\pi\rho_M\Omega_M\bar\Omega_M\tilde{b}_{1l}=0
      \label{a6d}\\
&(E + E_{l} + E_{l'} - E_1 - E_2 -U_{12}+
i\Gamma_M/2+ i\bar\Gamma_M/2\nonumber\\
&~~~~~~~~~~~~~~~+ i\Gamma_R/2) \tilde{b}_{12ll'}-
\Omega_{l'} \tilde{b}_{2l}+
\Omega_{l} \tilde{b}_{2l'}=0,
\label{a6e}\\
&~~~~~~~~~~~~~~~~~~~~~~\cdots
\nonumber
\end{align}
\end{subequations}
where $\Gamma_M = 2\pi\rho_M|\Omega_M|^2$ and
$\bar\Gamma_M = 2\pi\rho_M|\bar\Omega_M|^2$ are
the partial widths of the levels $E_1$ and $E_2$, respectively due
to coupling to the ballistic channel with the density of states $\rho_M$, and $\Gamma_{L,R} = 2\pi\rho_{L,R}|\Omega_{R,L}|^2$ are partial widths
due to coupling to the emitter and collector.
The Coulomb interdot repulsion $U_{12}$ just shifts the energy of the corresponding dot, as in Eq.~(\ref{a6e}). In the case of Coulomb blockade, $U_{12}\to\infty$, the corresponding amplitude vanishes (c.f. with  Eqs.~(\ref{c3})).

The density matrix elements, Eqs.~(\ref{a7}), are directly related
to the amplitudes $\tilde b(E)$ through the inverse Laplace transform, Eqs.~(\ref{invlap}), (\ref{invlapp}). Using this equation one can transform  Eqs. (\ref{a6}) for the amplitudes $\tilde b(E)$
into differential equations for the probabilities $\sigma^{(k,n)}(t)$.
Consider, for instance, the term
$\sigma_{bb}^{(0,0)}(t)=\sum_l |b_{1l}(t)|^2$,
Eq.~(\ref{a7b}). Multiplying Eq.~(\ref{a6b}) by $\tilde{b}^*_{1l}(E')$
and then subtracting  the complex conjugated
equation with the interchange
$E\leftrightarrow E'$ we obtain
\begin{align}
&\int\sum_l\left\{(E'-E
-i\Gamma_M)\tilde b_{1l}(E)\tilde b^*_{1l}(E')-\Omega_l\big[\tilde b_{1l}(E)\tilde b^*_0(E')\right.\nonumber\\
&\left.~~~~~~~~~~-\tilde b^*_{1l}(E')\tilde b_0(E)\big]
-i\pi\rho_M\Omega_M\bar\Omega_M
[\tilde b_{1l}(E)\tilde b^*_{2l}(E')
\right.\nonumber\\
&\left.~~~~~~~~~~
+\tilde b^*_{1l}(E')\tilde b_{2l}(E)]
\right\}e^{i(E'-E)t}\frac{dEdE'}{4\pi^2}=0.
\label{a10}
\end{align}
Substituting
\begin{equation}
\tilde b_{1l}(E)=\frac{\Omega_l\tilde b_0(E)-
i\pi\rho_M\Omega_M\bar\Omega_M\tilde b_{2l}(E)}
{E+E_l-E_1+i\Gamma_M/2}
\label{a11}
\end{equation}
into Eq.~(\ref{a10}) we can carry out the $E,E'$-integrations thus
obtaining
\begin{align}
&\dot{\sigma}^{(0,0)}_{bb} = - \Gamma_M \sigma_{bb}^{(0,0)}
+\Gamma_L \sigma_{aa}^{(0,0)}\nonumber\\
&~~~~~~~~~~~~~~~~~~~~
-\pi\rho_M\Omega_M\bar\Omega_M(\sigma^{(0,0)}_{bc}+\sigma^{(0,0)}_{cb})
\label{a12},
\end{align}

Applying the same procedure to each of equations (\ref{a6c}), we obtain the following Bloch-type rate equations for the matrix
elements of the density-submatrix $\sigma (t)$:
\begin{subequations}
\label{a13}
\begin{align}
&\dot{\sigma}^{(k,n)}_{aa} = - \Gamma_L \sigma_{aa}^{(k,n)}
+\Gamma_M\sigma_{bb}^{(k-1,n)}+\bar\Gamma_M\sigma_{cc}^{(k-1,n)}
\nonumber\\
&
+\Gamma_R\sigma_{cc}^{(k,n-1)}
+2\pi\rho_M\Omega_M\bar\Omega_M(\sigma^{(k-1,n)}_{bc}+\sigma^{(k-1,n)}_{cb})
\label{a13a}\\
&\dot{\sigma}^{(k,n)}_{bb} = - \Gamma_M \sigma_{bb}^{(k,n)}
+\Gamma_L\sigma_{aa}^{(k,n)}+\bar\Gamma_M\sigma_{dd}^{(k-1,n)}
\nonumber\\
&+\Gamma_R\sigma_{dd}^{(k,n-1)}
-\pi\rho_M\Omega_M\bar\Omega_M(\sigma^{(k,n)}_{bc}+\sigma^{(k,n)}_{cb})
\label{a13b}\\
&\dot{\sigma}^{(k,n)}_{cc} = -(\Gamma_L +\bar\Gamma_M +\Gamma_R)
\sigma_{cc}^{(k,n)}
+\Gamma_M\sigma_{dd}^{(k-1,n)}\nonumber\\
&~~~~~~~~~~~~~~~~~~~~
-\pi\rho_M\Omega_M\bar\Omega_M(\sigma^{(k,n)}_{bc}+\sigma^{(k,n)}_{cb})
\label{a13c}\\
&\dot{\sigma}^{(k,n)}_{dd} = -(\Gamma_M +\bar\Gamma_M +\Gamma_R)
\sigma_{dd}^{(k,n)}+\Gamma_L\sigma_{cc}^{(k,n)}
\label{a13d}\\
&\dot{\sigma}^{(k,n)}_{bc} = i(E_2-E_1)\sigma^{(k,n)}_{bc}
-\pi\rho_M\Omega_M\bar\Omega_M(\sigma^{(k,n)}_{bb}+\sigma^{(k,n)}_{cc})
\nonumber\\
&
-2\pi\rho_M\Omega_M\bar\Omega_M\sigma_{dd}^{(k-1,n)}
-\frac{1}{2}(\Gamma_L+\Gamma_M +\bar\Gamma_M +\Gamma_R)
\sigma_{bc}^{(k,n)}
\label{a13e}
\end{align}
\end{subequations}

Equations (\ref{a13}) have clear physical interpretation. Consider
for instance Eq.~(\ref{a13a}) for the probability rate of finding
the system in the state $a$ with
$k$ electrons in the ballistic channel and $n$ electrons
in the right reservoir (Fig.~\ref{fig4}a). This state decays with the rate $\Gamma_L$
into the state $b$ (Fig.~\ref{fig4}b) whenever an electron enters the first
dot from the left reservoir. This process is
described by the first term in Eq.~(\ref{a13a}). On the other hand,
the states $b$ and $c$ (Figs.~\ref{fig4}b,c) with $k-1$ electrons
in the ballistic channel
decay into the state $a$ with $k$ electrons in the ballistic channel.
It takes place due to one-electron tunneling from the quantum dots
into the ballistic channel with the rates $\Gamma_M$ and
$\bar\Gamma_M$ respectively. This process is described by
the second and the third terms in Eq.~(\ref{a13a}). Also
the state $c$ (Fig.~\ref{fig4}c) with $n-1$ electrons in the
right reservoir can decay into the state $a$ due to tunneling
to the right reservoir with the rate $\Gamma_R$
(the fourth term in Eq.~(\ref{a13a})).  The last term in this
equation describes the decay of coherent superposition of the
states $b$ and $c$ into the state $a$. It takes place due to single
electron tunneling from the first and the second dots into the same state
of the ballistic channel with the amplitudes $\Omega_M$ and $\bar\Omega_M$,
respectively. Obviously, this process has no classical analogy,
since classical particle cannot simultaneously occupy two dots.

Equations (\ref{a13b}), (\ref{a13c}) and (\ref{a13d}) describe the
probability rate of finding the system in the states
where one of the dots or both dots are occupied. In the first case
an electron can jump into unoccupied dot via continuum states of the
ballistic channel. As a result, the states $b$ and $c$ can decay into
linear superposition of the states $b$ and $c$. This process is
described by the last terms in Eqs.~(\ref{a13b}), (\ref{a13c}).
Obviously, if the both dots are occupied, such a process cannot
take a place. Therefore
$\sigma_{dd}$ is not coupled with the nondiagonal density-matrix
elements, Eq.~(\ref{a13d}). The last equation, (\ref{a13e}) describes
the time-dependence of the nondiagonal density matrix element.
It has the same interpretation as all previous equations.

Equations (\ref{a13}) for the reduced density matrix $\sigma(t)$
were derived starting from the wave function $|\Psi (t)\rangle$,
Eq.~(\ref{a2}), where all the levels of
the reservoirs are occupied up to the corresponding Fermi energies.
In fact, at finite temperature, the system is not initially in a pure
state, but in incoherent superposition of different pure state, weighted by the corresponding Fermi function. The final answer therefore should be averaged over this distribution. In our case, however, it is not relevant, since we consider the large bias limit. Then energy levels $E_{1,2}$ of the dots are far away from the Fermi-levels, such that ($T\ll \mu_L-E_{1,2},\; E_{1,2}-\mu_R$). Thus, the reservoir levels
that carry the current ($|E_{l}-E_{1,2}|\lesssim \Gamma$)
are deeply inside the Fermi sea, so they can be considered as fully occupied. Hence, one would arrive to the same Eqs.~(\ref{a13}).

Using Eqs.~(\ref{a13}) one finds for the total current,
Eq.~(\ref{a8})
\begin{equation}
I(t)=\Gamma_R[\sigma_{cc}(t)+\sigma_{dd}(t)],
\label{a14}
\end{equation}
where $\sigma_{ii}=\sum_{k,n}\sigma^{(k,n)}_{ii}$ are the total
``probabilities''. We can easily understand this result
by taken into account that  $\sigma_{cc}+\sigma_{dd}$ is the
total probability for occupation of the second dot and
$\Gamma_R$ is the rate of electron transitions from this
dot to the (adjacent) right reservoir.

In order to find differential equations for $\sigma_{ij}$ we sum
over $k,n$ in Eqs.~(\ref{a13}). Then we obtain the following
Bloch-type equations, which describe the time-dependence of the
density-matrix for separated dots \cite{xq3}
\begin{subequations}
\label{a15}
\begin{align}
&\dot{\sigma}_{aa} = - \Gamma_L \sigma_{aa}
+\Gamma_M\sigma_{bb}+\bar\Gamma_M\sigma_{cc}
+\Gamma_R\sigma_{cc}\nonumber\\
&~~~~~~~~~~~~~~~~~~~~~~~~~~~
+2\pi\rho_M\Omega_M\bar\Omega_M(\sigma_{bc}+\sigma_{cb})
\label{a15a}\\
&\dot{\sigma}_{bb} = - \Gamma_M \sigma_{bb}
+\Gamma_L\sigma_{aa}+\bar\Gamma_M\sigma_{dd}
+\Gamma_R\sigma_{dd}\nonumber\\
&~~~~~~~~~~~~~~~~~~~~~~~~~~~~~
-\pi\rho_M\Omega_M\bar\Omega_M(\sigma_{bc}+\sigma_{cb})
\label{a15b}\\
&\dot{\sigma}_{cc} = -(\Gamma_L +\bar\Gamma_M +\Gamma_R)
\sigma_{cc}
+\Gamma_M\sigma_{dd}\nonumber\\
&~~~~~~~~~~~~~~~~~~~~~~~~~~~~~
-\pi\rho_M\Omega_M\bar\Omega_M(\sigma_{bc}+\sigma_{cb})
\label{a15c}\\
&\dot{\sigma}_{dd} = -(\Gamma_M +\bar\Gamma_M +\Gamma_R)
\sigma_{dd}+\Gamma_L\sigma_{cc}
\label{a15d}\\
&\dot{\sigma}_{bc} = i(E_2-E_1)\sigma_{bc}
-\pi\rho_M\Omega_M\bar\Omega_M(\sigma_{bb}+\sigma_{cc})\nonumber\\
&~~~~~~~~~~~~~~~~~~~~~~
-2\pi\rho_M\Omega_M\bar\Omega_M\sigma_{dd}-\frac{1}{2}
\Gamma_{tot}\sigma_{bc},
\label{a15e}
\end{align}
\end{subequations}
where $\Gamma_{tot}=\Gamma_L+\Gamma_M +\bar\Gamma_M +\Gamma_R$. Note that $2\pi\rho_M\Omega_M\bar\Omega_M=\pm\sqrt{\Gamma_M\bar\Gamma_M}$, since the amplitudes $\Omega_M$, $\bar\Omega_M$ can be of the opposite signs due to different parity of the dots localized states \cite{parity}.

The stationary (dc) current $I=I(t\to\infty )$
Eq.~(\ref{a14}) can be easily obtained from Eqs.~(\ref{a15}) by taken
into account that $\dot\sigma_{ij}\to 0$ for $t\to\infty$. As a result,
Eqs.~(\ref{a15}) turn into a system of linear algebraic equations,
supplemented by a probability conservation condition
$\sigma_{aa}+\sigma_{bb}+\sigma_{cc}+\sigma_{dd}=1$. Consider, for example,
the case of the same of $\Gamma_L=\Gamma_M=\bar\Gamma_M\equiv \Gamma$. Solving Eqs.~(\ref{a14}), (\ref{a15}) one finds for dc current \cite{xq3}
\begin{equation}
I=\frac{\Gamma^2\Gamma_R^{}(3\Gamma+\Gamma_R)}{8\epsilon^2(\Gamma+\Gamma_R)
+(3\Gamma+\Gamma_R)^2(\Gamma+2\Gamma_R)},
\label{a16}
\end{equation}
where $\epsilon =E_1-E_2$.

Similar to the coupled-dot case, Eq.~(\ref{b11}), dc current
in separated dots displays the Lorentzian shape resonance as
a function of $\epsilon$ and the same peculiar dependence on the
coupling with the collector. Indeed, contrary to expectations, the current vanishes when $\Gamma_R\to\infty$. The latter manifests the quantum-coherence
effects in double-dot and separated dot systems \cite{xq3}.

\section{General case}

The (number resolved) rate equations (\ref{a13}), describing electron
transport in separated dots, can be extended to any multi-dot system.
By applying the same technique of integrating out the reservoir
states discussed above, we arrive to the rate equations for the density-matrix $\sigma_{\alpha\beta}^{(n,m,\ldots)}$
of the multi-dot system, where $n,m,\ldots$ denote the number of electrons, arriving to corresponding reservoirs. Tracing over $n,m,\ldots$, these equations can be written in a general form as \cite{xq3,gm}
\begin{align}
\dot\sigma_{\alpha\beta} &=  i(E_\beta - E_\alpha) \sigma_{\alpha\beta} +
i\Big (\sum_{\gamma}\sigma_{\alpha\gamma}
\tilde\Omega_{\gamma\to\beta}
-\sum_{\gamma}\tilde\Omega_{\alpha\to\gamma}
\sigma_{\gamma\beta}\Big)\nonumber\\
&-{\sum_{\gamma,\delta}}\mathcal{P}_2
\pi\rho(\sigma_{\alpha\gamma}\Omega_{\gamma\to\delta}\Omega_{\delta\to\beta}
+\sigma_{\gamma\beta}\Omega_{\gamma\to\delta}\Omega_{\delta\to\alpha})\nonumber\\
&+\sum_{\gamma,\delta}\mathcal{P}_2
2\pi\rho\,\Omega_{\gamma\to\alpha}\Omega_{\delta\to\beta}\,
\sigma_{\gamma\delta}\, ,
\label{d1}
\end{align}
where $|\alpha\rangle,\,  |\beta\rangle,\ldots$ denote all {\em discrete} states of
the multi-dot system in the occupation number representation, and
$\Omega_{\alpha\to\beta}$ denotes one-electron
hopping amplitude that generates $\alpha\to\beta$-transition. We distinguish
between the amplitudes $\tilde\Omega$ and $\Omega$ of
one-electron hopping among isolated states and among isolated
and continuum states, respectively. The latter transitions are
of the second order in the hopping amplitude $\sim\Omega^2$.
These transition are produced by two consecutive hoppings of an electron across continuum states with the density of states $\rho$. The first of these terms arises from ``loss'' processes and the second  from ``gain'' processes (borrowing the terminology of the classical Boltzmann equation). $\mathcal{P}_2$ is the Pauli factor: $\mathcal{P}_2=-1$ in  transitions involving two electrons, +1 otherwise.

Applying these rules, it is rather easy to verify that Eqs.~(\ref{d1}) coincide with
Eqs.~({\ref{a15}) for $\alpha,\beta,\ldots = \{a,b,c,d\}$,
which are the states of the separated dot system, shown in Fig.~\ref{fig4}. In addition,
Eqs.~(\ref{d1}) have the same form for the number-resolved density matrix, $\sigma_{\alpha\beta}^{(n,m,\ldots)}$. One only needs to take into account that the number of electron indicated in each of the term in the rhs of this equation is the same as in the lhs, if the electron returns to the same reservoir, or it is less by one, if a new electron arrives to the another reservoir, (see Eqs.~(\ref{c4}),(\ref{b8}),(\ref{a13})).

It is easy to realize that Eq.~(\ref{d1}) has precisely a form of the Lindblad equation \cite{lind}, which for the $N$-dimensional system reads
\begin{align}
\dot\sigma&=i[H_N,\sigma]-\sum_{j=1}^{N^2-1} \gamma_j\big(\sigma A_j^\dagger A_j^{}+A_j^\dagger A_j^{}\sigma\big)\nonumber\\
&~~~~~~~~~~~~~~~~~~~~~~~~
+\sum_{j=1}^{N^2-1} 2\gamma_jA_j^{}\sigma A_j^\dagger
\end{align}
where $H_N$ is the system Hamiltonian, $A_j$ is the Lindblad operator and $\gamma_j>0$. However, one should emphasize, that the necessary conditions for derivation of Eq.~(\ref{d1}) are the Markovian environment and reservoirs (energy independent spectral density function) and large bias limit.

Applications of Eq.~(\ref{d1}) for spin-polarized and unpolarized current through two-levels dots can be found in \cite{gmb}.

\section{Description of measurement by rate equations}

\subsection{Ballistic point-contact detector}
Consider the measurement of electron occupation
of a semiconductor quantum dot by
means of a separate measuring circuit in close proximity\cite{pep, buks3}.
A ballistic one-dimensional point-contact is used as a ``detector'' that
resistance is very sensitive to the electrostatic field generated
by an electron occupying the measured quantum dot. Such a set up is shown
schematically in Fig.~\ref{mfig1}, where the detector is represented by
a barrier, connected with two reservoirs at the chemical potentials
$\mu_L$ and $\mu_R$ respectively. The transmission probability
of the barrier varies from $T$ to $T'$, depending on whether or not
the quantum dot is occupied by an electron, Fig.~\ref{mfig1}(a,b).

Initially all the levels in the reservoirs are filled up to the
corresponding Fermi energies and the quantum dot is empty.
(For simplicity we consider the reservoirs at zero temperature).
The time-evolution of the entire system can
be described by the following master (rate) equations. which we derive in next subsections
\begin{subequations}
\label{am1}
\begin{align}
\dot\sigma_{aa}^{(m,n)}&=-(\Gamma_L+D)\sigma_{aa}^{(m,n)}
+\Gamma_R\sigma_{bb}^{(m-1,n)}+D\sigma_{aa}^{(m,n-1)}
\label{am1a}\\
\dot\sigma_{bb}^{(m,n)}&=-(\Gamma_R+D')\sigma_{bb}^{(m,n)}
+\Gamma_L\sigma_{aa}^{(m,n)}+D'\sigma_{bb}^{(m,n-1)}
\label{am1b}
\end{align}
\end{subequations}
where $\sigma_{aa}^{m,n}(t)$ and $\sigma_{bb}^{m,n}(t)$ are probabilities
of finding the entire system in the states $|a\rangle$ and
$|b\rangle$ corresponding to empty or occupied dot Fig.~\ref{mfig1}(a,b), and
$m$ and $n$ are the number of electrons penetrated to the right reservoirs of the measured system and the detector, respectively.
$\Gamma_{L,R}$ are the transition rates for an
electron tunneling from the left reservoir to the dot
and from the dot to the right reservoir respectively,
and $D=T(\mu_L-\mu_R)/2\pi$ is the rate of electron hopping
from the right to the left reservoir through the point-contact
(the Landauer formula).
\begin{figure}[tbh]
\includegraphics[width=8cm]{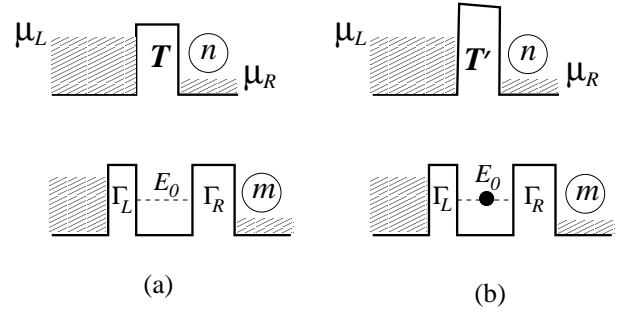}
\caption{Ballistic point-contact near quantum-dot. $\Gamma_{L,R}$
are the corresponding tunneling rates. The penetration coefficient
of the point-contact is $T$ for the empty dot (a) and $T'$ for the occupied
dot (b). The indices $m$ and $n$ denote the number of electrons
penetrating to the right reservoirs at time $t$.}
\label{mfig1}
\end{figure}

The accumulated charge in the right reservoirs of the detector ($d$)
and of the measured system ($s$) is given by
\begin{subequations}
\label{am2}
\begin{align}
Q_d(t)&=\sum_{m,n}n[\sigma_{aa}^{(m,n)}(t)+\sigma_{bb}^{(m,n)}(t)]
\label{am2a}\\
Q_s(t)&=\sum_{m,n}m[\sigma_{aa}^{(m,n)}(t)+\sigma_{bb}^{(m,n)}(t)]
\label{am2b}
\end{align}
\end{subequations}
The currents flowing in the detector and in the measured system
are $I_d(t)=\dot Q_d(t)$ and $I_s(t)=\dot Q_s(t)$.
Using Eqs.~(\ref{am1}) and (\ref{am2}) we obtain
\begin{subequations}
\label{aam3}
\begin{align}
I_d(t) &=\sum_{m,n}n[\dot\sigma_{aa}^{(m,n)}(t)
+\dot\sigma_{bb}^{(m,n)}(t)]=
D\sigma_{aa}(t)+D'\sigma_{bb}(t)
\label{aam3a}\\
I_s(t) &=\sum_{(m,n)}m[\dot\sigma_{aa}^{(m,n)}(t)
+\dot\sigma_{bb}^{(m,n)}(t)]=
\Gamma_R\sigma_{bb}(t)\ ,
\label{aam3b}
\end{align}
\end{subequations}
where
$\sigma_{aa}\equiv\sum_{m,n}\sigma_{aa}^{(m,n)}$ and
$\sigma_{bb}\equiv\sum_{m,n}\sigma_{bb}^{(m,n)}$ are the total
probabilities of finding the dot empty or occupied. Obviously
$\sigma_{aa}(t)=1-\bar\sigma (t)$, where $\bar\sigma (t)
\equiv \sigma_{bb} (t)$.
Performing the summation over $m,n$ in Eqs.~(\ref{am1}) we obtain the
following rate equation for the quantum dot occupation probability
\begin{equation}
\dot{\bar\sigma}(t)=\Gamma_L -(\Gamma_L+\Gamma_R)\bar\sigma (t)\ .
\label{am4}
\end{equation}

If the point-contact and the quantum dot are decoupled,
the detector current is $I_d^{(0)}=D$.
Hence, the occupation of the quantum dot can be measured
through the variation of the detector current $\Delta I_d = I_d^{(0)}-I_d$.
One readily obtains from Eq.~(\ref{aam3b}) that
\begin{equation}
\Delta I_d(t)=\frac{\Delta T\, V_d}{2\pi}\bar\sigma (t),
\label{am8}
\end{equation}
where $V_d=\mu_L-\mu_R$ is the voltage bias, and $\Delta T=
T-T'$. Thus, the point contact is indeed the measurement device.
In fact, Eq.~(\ref{am8}) is a self-evident one. Indeed, the variation
of the point-contact current is $\Delta TV_d/2\pi$ and $\bar\sigma(t)$
is the probability for such a variation.

\subsection{Derivation of the rate equations for a point-contact detector}

We present here the microscopic derivation of the rate equations describing
electron transport through the point contact. The latter is considered as a barrier, separated two reservoirs (the emitter and the collector), Fig.~\ref{mfig1}. The system is described by the tunneling Hamiltonian
\begin{align}
H_{PC}=\sum_l E_la_l^\dagger a_l+\sum_r E_ra_r^\dagger a_r
+\sum_{l,r}\Omega_{lr}(a_l^\dagger a_r+H.c.),
\label{apm1}
\end{align}
where $a_l^\dagger (a_l)$ and $a_r^\dagger  (a_r)$ are the creation
(annihilation) operators in the left and the right
reservoirs, respectively, and $\Omega_{lr}$ is the hopping amplitude
between the states $E_l$ and $E_r$ in the right and the left reservoirs.
(We choose the the gauge where $\Omega_{lr}$ is real).

Consider all the levels in the emitter and the collector are initially filled up to the Fermi energies $\mu_L$ and $\mu_R$ respectively, Fig.~\ref{mfig1}. We call
it as the ``vacuum'' state, $|0\rangle$. The Hamiltonian Eq.~(\ref{apm1}) requires the vacuum state $|0\rangle$ to decay to continuum states
having the form: $a_{r}^{\dagger}a_{l}|0\rangle$ with an electron
in the collector continuum
and a hole in the emitter continuum;
$a_{r}^{\dagger}a_{r'}^{\dagger}a_{l}^{\dagger}a_{l'}|0\rangle$
with two electrons in the collector continuum and two holes in
the emitter continuum, and so on.
The many-body wave function describing this system can be written
in the occupation number representation as
\begin{align}
|\Psi (t)\rangle &= \left [ b_0(t) + \sum_{l,r} b_{lr}(t)a_r^{\dagger}a_l\right.\nonumber\\
&\left.~~~~~
+\sum_{l<l',r<r'} b_{ll'rr'}(t)a_r^{\dagger}a_{r'}^{\dagger}a_la_{l'}
+\cdots\right ]|0\rangle,
\label{apm2}
\end{align}
where $b(t)$ are the time-dependent probability amplitudes to
find the system in the corresponding states with the initial
condition $b_0(0)=1$, and all the other $b(0)$'s being zeros.
Substituting Eq.~(\ref{apm2}) into the Shr\"odinger equation
$i|\dot\Psi (t)\rangle ={\cal H}_{PC}|\Psi (t)\rangle$
and performing the Laplace transform $b(t)\to\tilde b(E)$, Eq.~(\ref{lap}), we obtain an infinite set of the coupled equations for the amplitudes $\tilde b(E)$:
\begin{subequations}
\label{apm4}
\begin{align}
& E \tilde{b}_{0}(E) - \sum_{l,r} \Omega_{lr}\tilde{b}_{lr}(E)=i
\label{apm4a}\\
&(E + E_{l} - E_r) \tilde{b}_{lr}(E) - \Omega_{lr}\tilde{b}_0(E)~\nonumber\\
&~~~~~~~~~~~~~~~~~~~~~~~~~~~
-\sum_{l',r'}\Omega_{l'r'}\tilde{b}_{ll'rr'}(E)=0
\label{apm4b}\\
&(E + E_{l}+E_{l'} - E_r-E_{r'}) \tilde{b}_{ll'rr'}(E)
- \Omega_{l'r'}\tilde{b}_{lr}(E)\nonumber\\
&+\Omega_{lr}\tilde{b}_{l'r'}(E)
-\sum_{l'',r''}\Omega_{l''r''}\tilde{b}_{ll'l''rr'r''}(E)=0
\label{apm4c}\\
&~~~~~~~~~~~~~~~~~~~~~~~~~~~~~~~~\cdots
\nonumber
\end{align}
\end{subequations}
Eqs. (\ref{apm4}) can be substantially simplified by replacing
the amplitude $\tilde b$ in
the term $\sum\Omega\tilde b$ of each of the equations  by
its expression obtained from the subsequent equation (c.f. Eqs.~({\ref{exam}), (\ref{a5})). For example,
substituting $\tilde{b}_{lr}(E)$ from Eq.~(\ref{apm4b}) into Eq.~(\ref{apm4a}),
one obtains
\begin{align}
&\left [ E - \sum_{l,r}\frac{\Omega^2}{E + E_{l} - E_r}
    \right ] \tilde{b}_{0}(E)\nonumber\\
&~~~~~~~~~~~~~~~~- \sum_{ll',rr'}
    \frac{\Omega^2}{E + E_{l} - E_r}\tilde{b}_{ll'rr'}(E)=i,
\label{apm5}
\end{align}
where we assumed that the hopping amplitudes
are weakly dependent functions on the energies
$\Omega_{lr}\equiv\Omega (E_l,E_r)=\Omega$ (wide-band limit).
Since the states in the reservoirs are very dense (continuum),
one can replace the sums over $l$ and $r$ by integrals, for instance
$\sum_{l,r}\;\rightarrow\;\int \rho_{L}(E_{l})\rho_{R}(E_{r})\,dE_{l}dE_r\:$,
where $\rho_{L,R}$ are the density of states in the emitter and collector.
Then the first sum in Eq.~(\ref{apm5}) becomes an
integral which can be split into a sum of the singular and principal value
parts. The singular part yields $i\pi\Omega^2\rho_L\rho_R V_d$, whereas the principal value part can be neglected in the large bias limit, Eqs.~(\ref{exam1}), (\ref{exam2}). The second sum in Eq.~({\ref{apm5})
can be neglected either. Indeed, by replacing
$\tilde{b}_{ll'rr'}(E)\equiv \tilde{b} (E,E_l,E_{l'},E_r,E_{r'})$ and
the sums by the integrals we find that the integrand
has the poles on the same sides of the integration
contours. It implies that the corresponding integral vanishes (c.f. with Eqs.~(\ref{exam3}), (\ref{eexam5})).

Applying analogous considerations to the other equations of the
system (\ref{apm4}), we finally arrive to the following set of equations:
\begin{subequations}
\label{apm6}
\begin{align}
& (E + iD/2) \tilde{b}_{0}=i
\label{apm6a}\\
& (E + E_{l} - E_r + iD/2) \tilde{b}_{lr}
      - \Omega\tilde{b}_{0}=0
\label{apm6b}\\
& (E + E_{l}+ E_{l'} - E_{r} - E_{r'} + iD/2) \tilde{b}_{ll'rr'} -
      \Omega\tilde{b}_{lr}\nonumber\\
      &~~~~~~~~~~~~~~~~~~~~~~~~~~~~~~~~~~~~~~~~
            +\Omega \tilde{b}_{l'r'}=0,
\label{apm6c}\\
&~~~~~~~~~~~~~~~~~~~~~~~~~~~~~~~~\cdots
\nonumber
\end{align}
\end{subequations}
where $D=2\pi\Omega^2\rho_L\rho_R V_d$.

The charge accumulated in the collector at time $t$ is
\begin{equation}
N_R(t) =\langle\Psi (t)|\sum_r a_r^\dagger a_r|\Psi (t)\rangle=
\sum_nn\sigma^{(n)}(t),
\label{apm7}
\end{equation}
where
\begin{align}
&\sigma^{(0)}(t)=|b_0(t)|^2,~~~~\sigma^{(1)}(t)=\sum_{l,r}|b_{lr}(t)|^2,\nonumber\\
&
\sigma^{(2)}(t)=\sum_{ll',rr'}|b_{ll'rr'}(t)|^2,\; \cdots
\label{apm8}
\end{align}
are the probabilities to find $n$ electrons in the collector.
These probabilities are directly related
to the amplitudes $\tilde b(E)$ through the inverse Laplace transform,
\begin{equation}
\sigma^{(n)}(t)=
\sum_{l\ldots , r\ldots}
\int \frac{dEdE'}{4\pi^2}\tilde b_{l\cdots r\cdots}(E)
\tilde b^*_{l\cdots r\cdots}(E')e^{i(E'-E)t}
\label{apm9}
\end{equation}
Using Eq.~(\ref{apm9}) one can transform  Eqs.~(\ref{apm6}) into the rate
equations for $\sigma^{(n)}(t)$. We find
\begin{subequations}
\label{apm10}
\begin{align}
&\dot{\sigma}^{(0)}(t) = -D\sigma^{(0)}(t)
\label{apm10a}\\
&\dot{\sigma}^{(1)}(t) = D\sigma^{(0)}(t)-D\sigma^{(1)}(t)
\label{apm10b}\\
&\dot{\sigma}^{(2)}(t) = D\sigma^{(1)}(t)-D\sigma^{(2)}(t)
\label{apm10c}\\
&~~~~~~~~~~~~~~~~~~~~~~\cdots
\nonumber
\end{align}
\end{subequations}

The operator, which defines the current flowing in this system is
\begin{equation}
\hat I=i\left[ H_{PC},\sum_r a_r^\dagger a_r\right ]=
i\sum_{l,r}\Omega_{lr}(a^\dagger_la_r
-a^\dagger_ra_l)
\label{apm11}
\end{equation}
Using Eqs.~(\ref{apm2}), (\ref{apm10}) and  (\ref{apm11}) we find for
the current
\begin{equation}
I =\langle\Psi (t)|\hat I|\Psi (t)\rangle=
D\sum_n\sigma^{(n)}(t)=D.
\label{apm12}
\end{equation}

\subsection{Transmission coefficient of the point-contact.}

Now we relate the coupling $\Omega$ of the tunneling Hamiltonian (\ref{apm1}) to the transmission coefficient $T$ of the point-contact. The latter
determines probability of penetration of a single electron from the left to the right lead at $t\to\infty$. For this reason we consider single-electron motion from the emitter to collector.
\begin{figure}[tbh]
\includegraphics[width=8cm]{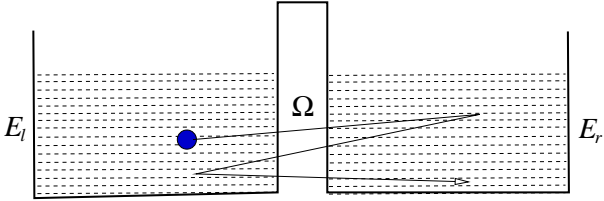}
\caption{Motion of a single electron between two reservoirs, described by the tunneling Hamiltonian~(\ref{apm1}).}
\label{mfig0}
\end{figure}

A single-electron wave-function in the basis of the reservoir states, ($|l\rangle$, $|r\rangle$) can be written as
\begin{align}
|\Psi(t)\rangle=\sum_l b_l^{(\bar l)}(t)|l\rangle+\sum_r b_r^{(\bar l)}(t)|r\rangle
\label{r0}
\end{align}
where the index $\bar l$ denotes the electron initial state, occupying a level $E_{\bar l}$ in the left reservoir. Substituting (\ref{r0}) in the Schr\"odinger equations, $i\partial_t|\Psi(t)\rangle= H_{PC}|\Psi(t)\rangle$, we find
\begin{align}
i{d\over dt}b_l^{(\bar l)}(t)&=E_l b_l^{(\bar l)}(t)+\sum_{r} \Omega\, b_{r}^{(\bar l)}(t)\nonumber\\
i{d\over dt}b_r^{(\bar l)}(t)&=E_r b_r^{(\bar l)}(t)+\sum_{l}\Omega\, b_{l}^{(\bar l)}(t)
\label{r1}
\end{align}

In order to solve these equations we apply the Laplace transform, $b(t)\to\tilde b(E)$, Eq.~(\ref{lap}). Then Eqs.~(\ref{r1}) become
\begin{align}
&(E-E_l)\tilde b_l^{(\bar l )}(E)-\sum_{r}\Omega\, \tilde b_{r}^{(\bar l)}(E)=i\delta_{l\bar l}^{}\nonumber\\
&(E-E_r)\tilde b_r^{(\bar l )}(E)-\sum_{l} \Omega\, \tilde b_{l}^{(\bar l)}(E)=0
\label{r2}
\end{align}
Let us introduce $\tilde B_L^{(\bar l)}(E)=\sum_l\tilde b_l^{(\bar l)}(E)$ and $\tilde B_R^{(\bar l)}(E)=\sum_r\tilde b_r^{(\bar l)}(E)$. Then Eqs.~(\ref{r2}) can be rewritten as
\begin{align}
\tilde B_L^{(\bar l)}(E)&=\Omega\,S_L\tilde B_R^{(\bar l)}(E)+{i\over E-E_{\bar l}}\nonumber\\
\tilde B_R^{(\bar l)}(E)&=\Omega\,S_R\tilde B_L^{(\bar l)}(E)
\label{r3}
\end{align}
where
\begin{align}
S_{L(R)}=\sum_{l(r)}{1\over E-E_{l(r)}}
\end{align}
In the continuous limit one obtains (c.f. Eqs.~(\ref{exam1}), (\ref{exam2}))
\begin{align}
S_{L(R)}\to\int\limits_{-\Lambda}^\Lambda {1\over E-E_{l(r)}}\rho_{L(R)}\, dE_{l(r)}^{}=
-i\pi \rho_{L(R)}^{}
\end{align}
where the cutoff $\Lambda\to\infty$. Then solving Eqs.~(\ref{r3}) one easily finds
\begin{align}
\tilde B_L^{(\bar l)}(E)={i\over (1+\pi^2\Omega^2\rho_L\rho_R)(E-E_{\bar l})}
\end{align}
and finally
\begin{align}
\tilde b_r^{(\bar l)}(E)={i\, \Omega\over (E-E_{\bar l})(E-E_r)(1+\pi^2\Omega^2\rho_L\rho_R)}
\label{r6}
\end{align}

The probability of finding the electron in the right lead at time $t$ is given by $P^{(\bar l)}_R(t)=\sum_r|b_r^{(\bar l)}(t)|^2$, where
\begin{align}
b_{r}^{(\bar l)}(t)=\int\limits_{-\infty}^\infty \tilde b_{r}^{(\bar l)}(E)e^{-iEt}{dE\over 2\pi}
\end{align}
Then using Eq.~(\ref{r6}) we obtain in the continuous limit
\begin{align}
&P^{(\bar l)}_R(t)=\sum_r\int\limits_{-\infty}^\infty \tilde b_r^{(\bar l)}(E)\tilde b_r^{(\bar l)*}(E')e^{i(E'-E)t}{dEdE'\over (2\pi)^2}\nonumber\\
&=\sum_r\int\limits_{-\infty}^\infty {\Omega^2e^{i(E'-E)t}dEdE'/(2\pi)^2\over (E-E_{\bar l})(E'-E_{\bar l})(E-E_{r})(E'-E_{r})(1+R)^2}\nonumber\\
&=-\int\limits_{-\infty}^\infty {i\Omega^2e^{i(E'-E)t}\rho_RdEdE'/(2\pi)\over (E-E_{\bar l})(E'-E_{\bar l})(E'-E)(1+R)^2}
\end{align}
where $R=\pi^2\Omega^2\rho_L\rho_R$.

Let us evaluate the electric current in the left reservoir, $I_R^{(\bar l)}(t)=dP_R^{(\bar l)}(t)/dt$. One finds
\begin{align}
I_R^{(\bar l)}(t)=
\int\limits_{-\infty}^\infty {(2\pi)\Omega^2e^{(E'-E)t}\rho_RdEdE'/(2\pi)^2\over (E-E_{\bar l})(E'-E_{\bar l})(1+R)^2}={2\pi \rho_R^{} \Omega^2\over (1+R)^2}
\end{align}
We consider the initial electron state as given by incoherent superposition of different states with a distribution $f_L(E_{\bar l})$. In the case we obtain
\begin{align}
I_R^{}(t)=
\int\limits_{-\infty}^\infty {2\pi\rho_L \rho_R^{} \Omega^2\over (1+R)^2}f_L(E_{\bar l})dE_{\bar l}
\end{align}
This in fact represents the Landauer formula
\begin{align}
I_R=\int T\,f_L(E_{\bar l}){dE_{\bar l}\over 2\pi}
\label{land1}
\end{align}
where
\begin{align}
T={(2\pi)^2\Omega^2\rho_L\rho_R\over (1+\pi^2\Omega^2\rho_L\rho_R)^2}\equiv{4R\over (1+R)^2}
\label{RT}
\end{align}
represents a relation between transmission coefficient $T$ and coupling $\Omega$ in the tunneling Hamiltonian.

Note that in the rate equations Eqs.~(\ref{apm10}), $D= RV_d/2\pi$, where $V_d$ is the bias. However, these equations were derived in the large bias limit, corresponding to $V_d\gg D$. The latter is equivalent to $R\ll 1$. In this case it follows from Eq.~(\ref{RT}) that $R=T/4$, in agreement with Eqs.~(\ref{am8}), (\ref{apm12}).

\section{Detection of electron oscillations in coupled-dots}

A well-known manifestation of quantum coherence is
the oscillation of a particle in a double-well (double-dot) potential.
The origin of these oscillations is the interference between the
probability amplitudes of finding a particle in different wells.
Hence, one can expect that the disclosure of a particle
(electron) in one of wells would generate the ``dephasing''
that eventually destroys these oscillations.

Let us investigate the mechanism of this process by taking for detector
a noninvasive point-contact. A possible set up is shown in Fig.~\ref{mfig2}.
\begin{figure}[tbh]
\includegraphics[width=8cm]{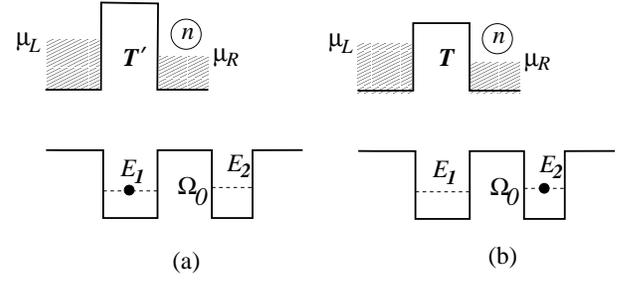}
\caption{Electron oscillations in the double-well. The penetration coefficient
of the point-contact varies from $T'$ to $T$ when an electron occupies the
left well (a) or the right well (b), respectively. The index $n$ denotes
the number of electrons accumulated in the collector at time $t$.}
\label{mfig2}
\end{figure}
We assume that the transmission probability of the point-contact is $T$
when an electron occupies the right well, and it is $T'$ when
an electron occupies the left well. Here $T'<T$ since
the right well is away from the point contact.

Now we apply the quantum-rate equations to the whole system.
However, in the distinction with the previous case, the electron
transitions in the measured system take place
between the {\em isolated} states inside the dots.
As a result the diagonal density-matrix elements are coupled with the
off-diagonal elements, so that the corresponding rate equations are
the Bloch-type equations Eqs.~(\ref{b8}).

We first start with the case of the double-well detached
from the point-contact detector.
The Bloch equations describing the time evolution of the electron
density-matrix $\sigma_{ij}$ have the following form
\begin{subequations}
\label{cm1}
\begin{align}
\dot\sigma_{aa}& = i\Omega_0 (\sigma_{ab}-\sigma_{ba})\;,
\label{cm1a}\\
\dot\sigma_{bb}& = i\Omega_0 (\sigma_{ba}-\sigma_{ab})\;,
\label{cm1b}\\
\dot\sigma_{ab}& = i\epsilon\sigma_{ab}+i\Omega_0(\sigma_{aa}
-\sigma_{bb}),
\label{cm1c}
\end{align}
\end{subequations}
where $\epsilon =E_2-E_1$ and $\Omega_0$
is the coupling between the left and the right wells.
Here $\sigma_{aa}(t)$ and $\sigma_{bb}(t)$ are
the probabilities of finding the electron in the left and the
right well respectively, and $\sigma_{ab}(t)=\sigma_{ba}^*(t)$
are the off-diagonal density-matrix elements (``coherences'')\cite{bloch}.

Solving these equations for the initial conditions and $\sigma_{aa}(0)=1$ and
$\sigma_{bb}(0)=\sigma_{ab}(0)=0$ we obtain
\begin{equation}
\sigma_{aa}(t)=\frac{\Omega_0^2\cos^2(\omega t)+\epsilon^2/4}{\Omega_0^2+
\epsilon^2/4},
\label{cm2}
\end{equation}
where $\omega =(\Omega_0^2+\epsilon^2/4)^{1/2}$. As expected the electron
initially localized in the first well oscillates between the wells
with the frequency $\omega$. Notice that the amplitude of these
oscillations is $\Omega_0^2/(\Omega_0^2+\epsilon^2/4)$.
Thus the electron remains localized in the first well if
the level displacement is large, $\epsilon \gg\Omega_0$.

Now we consider the electron oscillations in the presence of
the point contact detector, Fig.~\ref{mfig2}.
The corresponding Bloch equations for the entire system have the
following form, see Eq.~(\ref{d1}) (detailed microscopic derivation of these equations is given below)
\begin{subequations}
\label{cm3}
\begin{align}
&\dot\sigma_{aa}^{(n)}  =  -D'\sigma_{aa}^{(n)}+D'\sigma_{aa}^{(n-1)}
+i\Omega_0 (\sigma_{ab}^{(n)}-\sigma_{ba}^{(n)})\;,
\label{cm3a}\\
&\dot\sigma_{bb}^{(n)}  =  -D\sigma_{bb}^{(n)}
+D\sigma_{bb}^{(n-1)}-i\Omega_0 (\sigma_{ab}^{(n)}-\sigma_{ba}^{(n)})\;,
\label{c3b}\\
&\dot\sigma_{ab}^{(n)}  =  i\epsilon\sigma_{ab}^{(n)}+
i\Omega_0(\sigma_{aa}^{(n)}-\sigma_{bb}^{(n)})
-\frac{1}{2}(D'+D)\sigma_{ab}^{(n)}\nonumber\\
&~~~~~~~~~~~~~~~~~~~~~~~~~~~~~~~~~~~~
+(D\, D')^{1/2}\sigma_{ab}^{(n-1)}\, ,
\label{cm3c}
\end{align}
\end{subequations}
Here the index $n$ denotes the number of electrons arriving to the
collector at time $t$, and $D(D')$ is the transition rate of an electron
hopping from the left to the right detector reservoirs,
$D=T(\mu_L-\mu_R)/2\pi$, Eqs.~(\ref{apm6}). Notice that the presence of the detector
results in additional terms in the rate equations in comparison with
Eqs.~(\ref{cm1}). These terms are generated by transitions of
an electron from the left to the right detector reservoirs with
the rates $D$ and $D'$ respectively.
The equation for the non-diagonal
density-matrix elements $\sigma_{ab}^{(n)}$, Eq.~(\ref{cm3c}),
is different from the standard Bloch equations
due to the last term, which describes transition between different
coherences, $\sigma^{(n-1)}_{ab}$ and $\sigma^{(n)}_{ab}$.
This term appears in the Bloch equations for coherences
whenever the same hopping ($n-1\to n$) takes place in the {\em both}
states of the off-diagonal density-matrix element ($a$ and $b$), Eq.~(\ref{d1}).
The rate of such transitions is determined by
a product of the corresponding {\em amplitudes} ($T^{1/2}$ and
${T'}^{1/2}$).

It follows from Eqs.~(\ref{aam3}), (\ref{cm3})
that the variation of the point-contact
current $\Delta I_d(t)=
I^{(0)}-I_d(t)$ measures directly the charge in
the first dot. Indeed, one obtains for the detector current
\begin{equation}
I_d(t)=\sum_nn[\sigma_{aa}^{(n)}(t)+\sigma_{bb}^{(n)}(t)]=D'\sigma_{aa}(t)+
D\sigma_{bb}(t),
\label{ccm3}
\end{equation}
where $\sigma_{ij}=\sum_n\sigma_{ij}^{(n)}$. Therefore $\Delta I_d(t)$ is given
by Eq.~(\ref{am8}), where $\bar\sigma (t)\equiv \sigma_{aa}(t)$.

In order to determine the influence of the detector on the double-well
system we trace out the detector states in Eqs.~(\ref{cm3})
thus obtaining
\begin{subequations}
\label{cm4}
\begin{align}
\dot\sigma_{aa}& = i\Omega_0(\sigma_{ab}-\sigma_{ba})\;,
\label{cm4a}\\
\dot\sigma_{bb}& = i\Omega_0(\sigma_{ba}-\sigma_{ab})\;,
\label{cm4b}\\
\dot\sigma_{ab}& = i\epsilon\sigma_{ab}+i\Omega_0(\sigma_{aa}
-\sigma_{bb})\nonumber\\
&~~~~~~~~~~~~~~~~~~~~~~~~~~~~
-\frac{1}{2}(\sqrt{I}-\sqrt{I'})^2\sigma_{ab},
\label{cm4c}
\end{align}
\end{subequations}
where $\sigma_{ij}=\sum_n\sigma^{n}_{ij}(t)$, and $I,I'=D,D'$ are two values of the Point-Contact current, corresponding to the occupied right or left dot in Fig.~(\ref{mfig2}).

Equations~(\ref{cm4}) coincide with Eqs.~(\ref{cm1}), describing the electron oscillations without
detector, except for the last term in Eq.~(\ref{cm4c}). The latter
generates the exponential damping of the non-diagonal density-matrix
element with the ``dephasing'' rate
\begin{equation}
\Gamma_d=(\sqrt{I}-\sqrt{I'})^2=(\sqrt{T}-\sqrt{T'})^2\frac{V_d}{2\pi}\, ,
\label{cm5}
\end{equation}
It implies that $\sigma_{ab}\to 0$ for $t\to\infty$. We can check it by looking for the stationary solutions of Eqs.~(\ref{cm4}) in the limit $t\to\infty$. In this case $\dot\sigma_{ij}(t \to\infty )=0$ and Eqs.~(\ref{cm4}) become
linear algebraic equations, which can be easily solved.
One finds that the electron density-matrix becomes
the statistical mixture.
\begin{equation}
\sigma (t)=\left (\begin{array}{cc}
\sigma_{aa}(t)&\sigma_{ab}(t)\\
\sigma_{ba}(t)&\sigma_{bb}(t)\end{array}\right )
\to\left (\begin{array}{cc}
1/2&0\\0&1/2
\end{array}\right ) \;\; {\mbox{for}}\;\;\;  t\to\infty.
\label{cm6}
\end{equation}
Notice that the damping of the nondiagonal density matrix elements
is coming entirely from the possibility of disclosing the electron
in one of the wells. Indeed, if the detector does not
distinguish which of the wells is occupied,
i.e. $T=T'$, then $\Gamma_d=0$.

The Bloch equations (\ref{cm3}), (\ref{cm4}) display explicitly the
mechanism of dephasing during a noninvasive measurement,
i.e. that which does not distort
the energy levels of the measured system.
The dephasing appears in the reduced density matrix as the ``dissipative''
term in the nondiagonal density matrix elements only,
as a result of tracing out the detector variables. All other terms
related to the detector are canceled after tracing out the detector
variables.

\subsection{Continuous measurement and Zeno effect}

The most surprising phenomenon which displays Eq.~(\ref{cm6}) is that
the transition to the statistical mixture
takes place even for a large displacement of the energy levels,
$\epsilon\gg\Omega_0$, irrespectively of the initial conditions.
It means that an electron initially localized in one of the
wells would be always {\em delocalized} at $t\to\infty$.
It would happened even if the electron was initially localized
at the lower level. (Of course it does not
violate the energy conservation, since the double-well
is not isolated). Such a behavior
is not expectable because the amplitude of electron
oscillations is very small for large level displacement,
Eq.~(\ref{cm2}). Thus, the electron should
stay localized in one of the wells.
One could expect that the continuous observation of this electron
by a detector could only increase its localization.
It can be inferred from so called Zeno effect\cite{zeno}.
The latter tells us that repeated observation of the system
slow down transitions between quantum states due to the collapse
of the wave function into the observed state. Since in our case
the change of the detector current, $\Delta I_s(t)$
monitors $\bar\sigma (t)$ in the left well,
Eqs.~(\ref{am8}), (\ref{ccm3}), it represents the continuous measurement
of the charge in this well. Nevertheless the effects is just opposite --
the continuous measurement delocalizes the system (so-called, anti-Zeno effect \cite{anzeno}).

However, our results for small $t$ are in an agreement with the Zeno effect, even so
we have not explicitly implied the projection postulate.
For instance,  Fig.~\ref{mfig3}~a shows the time-dependence of the probability to
find an electron in the left dot, as obtained from the
solution of Eqs.~(\ref{cm4}) for the aligned levels ($\epsilon =0$), and
$\Gamma_d=0$ (dashed curve), $\Gamma_d=4\Omega_0$ (dot-dashed
curve) and  $\Gamma_d=16\Omega_0$ (solid curve). One finds that
for small $t$ the rate of transition from the left to the right well
decreases with the increase of $\Gamma_d$.
\begin{figure}[tbh]
\includegraphics[width=8cm]{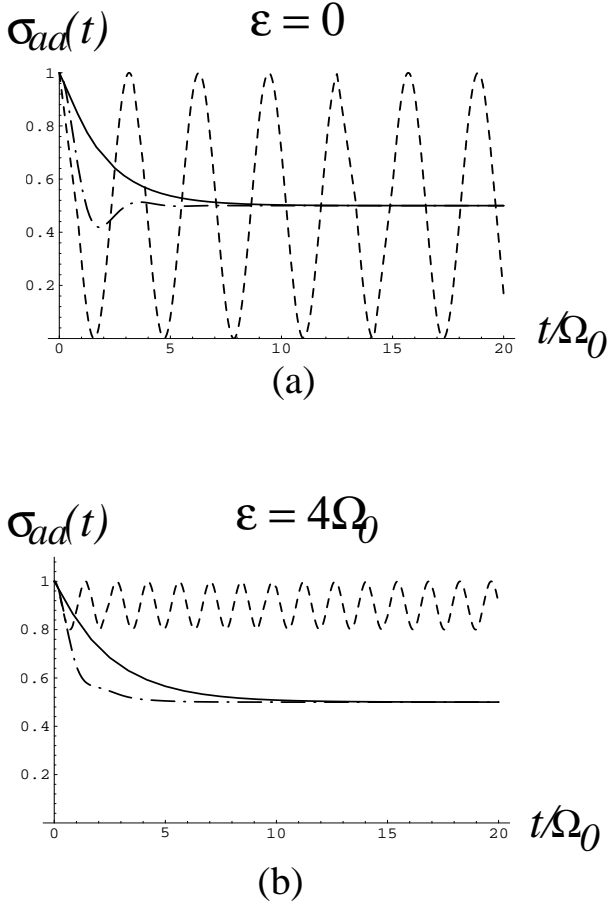}
\caption{The occupation of the first well as a function
of time, Eqs. (\ref{cm4}):
(a) the levels are aligned ($\epsilon =0$); (b) the
levels are displaced ($\epsilon =4\Omega_0$). The curves
correspond to different values of the dephasing rate:
$\Gamma_d=0$ (dashed),
$\Gamma_d=4\Omega_0$ (dot-dashed),  and $\Gamma_d=16\Omega_0$
(solid).}
\label{mfig3}
\end{figure}

The same slowing down of the transition rate for small enough $t$
we find for the disaligned  levels ($\epsilon =4\Omega_0$) in Fig.~\ref{mfig3}~b. However, with increase of $t$, the continuous measurement leads to the electron {\em delocalization} (the anti-Zeno effect \cite{anzeno,eg2}), whereas in the absence of detector an electron would stay localized in the left well (the dashed curve in Fig.~\ref{mfig3} b)

\subsection{Derivation of Master equations, describing double-dot under continuous measurement.}

Although the rate equations~(\ref{cm3}) is a particular case of general Master equations~(\ref{d1}), it is desirable to present a microscopic derivation of Eqs.~(\ref{cm3})  by using our wave-function approach.
We start with the many-body Schr\"odinger equation,
$i|\dot\Psi (t)\rangle =H|\Psi (t)\rangle$ for
the entire system. Here $H$ is
the total Hamiltonian, which can be written as
$H=H_{PC}+H_{DD}+H_{int}$.
Here $H_{PC}$ is the Hamiltonian for
the point-contact detector,
Eq.~(\ref{apm1}); $H_{DD}$ is the Hamiltonian for
the measured double-dot system,
\begin{equation}
H_{DD} = E_1 c_1^{\dagger}c_{1}+E_2 c_2^{\dagger}c_{2}+
              \Omega_0 (c_2^{\dagger}c_{1}+ c_1^{\dagger}c_{2})\, ,
\label{bpm1}
\end{equation}
and $H_{int}$ describes the interaction between the
detector and the measured system. Since the presence of an electron in the
left well results in an effective increase of the point-contact barrier
($\Omega_{lr}\to\Omega_{lr}+\delta\Omega_{lr}$), we can
represent the interaction term as
\begin{equation}
{\cal H}_{int}=\sum_{l,r}\delta\Omega_{lr}c_1^{\dagger}
c_1(a^{\dagger}_la_r+H.c.).
\label{bpm2}
\end{equation}

The many-body wave function for the entire system can be written as
\begin{align}
|\Psi (t)\rangle &= \left [ b_1(t)c_1^{\dagger}
+ \sum_{l,r} b_{1lr}(t)c_1^{\dagger}a_r^{\dagger}a_l\right.\nonumber\\
&\left.
+\sum_{l<l',r<r'} b_{1ll'rr'}(t)
c_1^{\dagger}a_r^{\dagger}a_{r'}^{\dagger}a_la_{l'}+\cdots\right.\nonumber\\
&\left.+b_2(t)c_2^{\dagger}
+ \sum_{l,r} b_{2lr}(t)c_2^{\dagger}a_r^{\dagger}a_l\right.\nonumber\\
&\left.+\sum_{l<l',r<r'} b_{2ll'rr'}(t)
c_2^{\dagger}a_r^{\dagger}a_{r'}^{\dagger}a_la_{l'}+\cdots
\right ]|0\rangle,
\label{bpm3}
\end{align}
where $b(t)$ are the probability amplitudes to find the entire
system in the states defined by the corresponding creation and
annihilation operators. Notice that Eq.~(\ref{bpm3}) has the same
form as Eq.~(\ref{apm2}), where only the probability amplitudes $b(t)$
acquire an additional index ('1' or '2') that denotes the well, occupied
by an electron. Proceeding in the same way as in Sec. VIII B,
we arrive to an infinite set of the coupled equations for the
amplitudes $\tilde b(E)$, which are the Laplace transform~(\ref{lap})
of the amplitudes $b(t)$
\begin{subequations}
\label{bpm4}
\begin{align}
& (E-E_1) \tilde{b}_{1}(E) - \Omega_0\tilde b_2(E)
-\sum_{l,r} \Omega'_{lr}\tilde{b}_{1lr}(E)=i
\label{bpm4a}\\
& (E-E_2) \tilde{b}_{2}(E)  - \Omega_0\tilde b_1(E)
- \sum_{l,r} \Omega_{lr}\tilde{b}_{2lr}(E)=0
\label{bpm4b}\\
&(E + E_{l}-E_1 - E_r) \tilde{b}_{1lr}(E) - \Omega'_{lr}\tilde{b}_1(E)
-\Omega_0\tilde b_{2lr}(E)\nonumber\\
&~~~~~~~~~~~~~~~~~~~~~~~~~~~
-\sum_{l',r'}\Omega_{l'r'}\tilde{b}_{1ll'rr'}(E)=0
\label{bpm4c}\\
&(E + E_{l}-E_2 - E_r) \tilde{b}_{2lr}(E) - \Omega_{lr}\tilde{b}_2(E)
-\Omega_0\tilde b_{1lr}(E)\nonumber\\
&~~~~~~~~~~~~~~~~~~~~~~~~~~~-\sum_{l',r'}\Omega_{l'r'}\tilde{b}_{2ll'rr'}(E)=0
\label{bp4md}\\
&~~~~~~~~~~~~~~~~~~~~~~~~\cdots
\nonumber
\end{align}
\end{subequations}
The same algebra as that used before allows us to
simplify these equations, which then become
\begin{subequations}
\label{bpm5}
\begin{align}
&(E -E_1+ iD'/2) \tilde{b}_{1}-\Omega_0\tilde b_2=i
\label{bpm5a}\\
&(E -E_2+ iD/2) \tilde{b}_{2}-\Omega_0\tilde b_1=0
\label{bpm5b}\\
&(E + E_{l} -E_1- E_r + iD'/2) \tilde{b}_{1lr}
      - \Omega'\tilde{b}_{1}\nonumber\\
&~~~~~~~~~~~~~~~~~~~~~~~~~~~~~~~~~~~~~~~~~~-\Omega_0\tilde b_{2lr}=0
\label{bpm5c}\\
&(E + E_{l} -E_2- E_r + iD/2) \tilde{b}_{2lr}
      - \Omega\tilde{b}_{2}\nonumber\\
&~~~~~~~~~~~~~~~~~~~~~~~~~~~~~~~~~~~~~~~~~~-\Omega_0\tilde b_{1lr}=0
\label{bpm5d}\\
&~~~~~~~~~~~~~~~~~~~~~~~~~~\cdots
\nonumber
\end{align}
\end{subequations}
where $D=TV_d/2\pi$.

Using the inverse Laplace transform (\ref{apm9}) we can transform
Eqs.~(\ref{bpm5}) into differential equations for the density-matrix
elements $\sigma^{(n)}_{ij}(t)$ ($i,j$=1,2)
\begin{align}
&\sigma^{(0)}_{ij}(t)=b_i(t)b^*_j(t),
~~~\sigma^{(1)}_{ij}(t)=\sum_{l,r}b_{ilr}(t)b^*_{jlr}(t),\nonumber\\
&\sigma^{(2)}_{ij}(t)=
\sum_{ll',rr'}b_{ill'rr'}(t)b^*_{jll'rr'}(t),\; \cdots\ ,
\label{bpm6}
\end{align}
where $n$ denotes the number of electrons
accumulated in the collector. Consider, for instance the off-diagonal
density-matrix element $\sigma^{(1)}_{12}(t)$. The corresponding
differential equation for this term can by obtained by
multiplying Eq.~(\ref{bpm5c}) by
$\tilde b^*_{2lr}(E')$ and subtracting the complex conjugated
Eq.~(\ref{bpm5d}) multiplied by $\tilde b_{1lr}(E)$. We then
obtain
\begin{align}
&\int\frac{dEdE'}{4\pi^2}\sum_{l,r}\left\{\left (E'-E-\epsilon
-i\frac{D+D'}{2}\right )\tilde b_{1lr}(E)\tilde b^*_{2lr}(E')
\right.\nonumber\\
&\left.
-\Big[\Omega \tilde b_{1lr}(E)\tilde b^*_2(E')
-\Omega' \tilde b^*_{2lr}(E')\tilde b_1(E)\Big]\right.\nonumber\\
&\left.-\Omega_0\Big[\tilde b_{1lr}(E)\tilde b^*_{1lr}(E')
-\tilde b^*_{2lr}(E')\tilde b_{2lr}(E)\Big]
\right\}e^{i(E'-E)t}=0.
\label{bpm7}
\end{align}
One easily finds that the first term
in this equation equals to
$-i\dot\sigma_{12}^{(1)}-[\epsilon+i(D+D')/2]\sigma_{12}^{(1)}$ and
the third term equals to $-\Omega_0(\sigma_{11}^{(1)}-
\sigma_{22}^{(1)})$. In order to evaluate the second term in Eq.~(\ref{bpm7})
we replace $\sum_{l,r}$ by the integrals and substitute
\begin{eqnarray}
\tilde b_{1lr}(E)=\frac{\Omega'\tilde b_1(E)+\Omega_0\tilde b_{2lr}(E)}
{E+E_l-E_1-E_r+iD'/2}\nonumber\\
\tilde b^*_{2lr}(E')=\frac{\Omega\tilde b^*_2(E')
+\Omega_0\tilde b^*_{1lr}(E')}
{E'+E_l-E_2-E_r-iD/2}
\end{eqnarray}
obtained from Eqs.~(\ref{bpm5c}), (\ref{bpm5d}),
into Eq.~(\ref{bpm7}). Then integrating over $E_l,E_r$ we find that
the second term in Eq.~(\ref{bpm7}) becomes
$2i\pi\Omega\Omega'\rho_L\rho_RV_d\sigma_{12}^{(0)}$. Thus
Eq.~(\ref{bpm7}) can be rewritten as
\begin{align}
&\dot\sigma_{12}^{(1)} = i\epsilon\sigma_{12}^{(1)}+
i\Omega_0(\sigma_{11}^{(1)}-\sigma_{22}^{(1)})
-\frac{1}{2}(D'+D)\sigma_{12}^{(1)}\nonumber\\
&~~~~~~~~~~~~~~~~~~~~~~~~~~~~~~~~~~~~~~
+(D\, D')^{1/2}\sigma_{12}^{(0)}\;.
\label{bpm9}
\end{align}
which coincides Eq.~(\ref{cm3c}) for $n=1$ and
$\sigma_{aa}\equiv\sigma_{11}$, $\sigma_{bb}\equiv\sigma_{22}$,
$\sigma_{ab}\equiv\sigma_{12}$. Applying the same procedure
to each of the equations (\ref{bpm5}) we arrive to Eqs.~(\ref{cm3}) for density matrix elements $\sigma_{ij}^{(n)}$.

\subsection{Measurement of resonant current through a double-dot.}

Measurement of decoherence rate, $\Gamma_d$, generated by a measurement device, is a very important issue in quantum computation. It can be done via direct monitoring of damping rate of the single-electron oscillations in coupled-dots,
Figs.~\ref{mfig2}, \ref{mfig3}. However, the same $\Gamma_d$ can be extracted from  the steady-state current, flowing through the double-dot. It is shown schematically in Fig.~\ref{mfig4}, where the coupled-dot is connected to two reservoirs (emitter and collector). For the sake of simplicity we assume strong
inner and inter-dot  Coulomb repulsion, so only one electron can occupy
this system. Then there are only three available states
of the coupled-dot system: the dots are empty ($a$),
the first dot is occupied ($b$)
and the second dot is occupied ($c$). Using Eq,~(\ref{d1}),
we write the following rate equations for the density matrix $\sigma^{m,n}_{ij}(t)$
describing the entire system (c.f. with Eqs.~(\ref{am1}), (\ref{cm3}))
\begin{subequations}
\label{bm5}
\begin{align}
&\dot\sigma_{aa}^{m,n}  =  -(\Gamma_L+D)\sigma_{aa}^{m,n}
+\Gamma_R\sigma_{cc}^{m-1,n}\nonumber\\
&~~~~~~~~~~~~~~~~~~~~~~~~~~~~~~~~~~~~~~~~~~
+D\sigma_{aa}^{m,n-1}\;,
\label{bm5a}\\
&\dot\sigma_{bb}^{m,n} =  -D'\sigma_{bb}^{m,n}+D'\sigma_{bb}^{m,n-1}
+\Gamma_L\sigma_{aa}^{m,n}\nonumber\\
&~~~~~~~~~~~~~~~~~~~~~~~~~~~~~~~~
+i\Omega_0 (\sigma_{bc}^{m,n}-\sigma_{cb}^{m,n})\;,
\label{bm5b}\\
&\dot\sigma_{cc}^{m,n}  =  -(\Gamma_R+D)\sigma_{cc}^{m,n}
+D\sigma_{cc}^{m,n-1}\nonumber\\
&~~~~~~~~~~~~~~~~~~~~~~~~~~~~~~~~-i\Omega_0 (\sigma_{bc}^{m,n}-\sigma_{cb}^{m,n})\;,
\label{bm5c}\\
&\dot\sigma_{bc}^{m,n} =  i\epsilon\sigma_{bc}^{m,n}+
i\Omega_0(\sigma_{bb}^{m,n}-\sigma_{cc}^{m,n})\nonumber\\
&~~
-\frac{1}{2}(\Gamma_R+D'+D)\sigma_{bc}^{m,n}
+(D\, D')^{1/2}\sigma_{bc}^{m,n-1}\, ,
\label{bm5d}
\end{align}
\end{subequations}
where the indices $n$ and $m$ denote the number of electrons arrived
at time $t$ to the upper and the lower collector reservoir, respectively.
Here  $\Gamma_{L}$, $\Gamma_{R}$ are the rates of electron transitions
from the left reservoir to the first dot and from the second dot
to the right reservoir, and $\Omega_0$ is the amplitude of hopping between
two dots.
\begin{figure}[tbh]
\includegraphics[width=5cm]{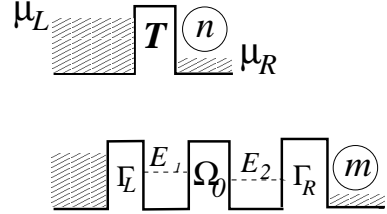}
\caption{Resonant tunneling through the double-dot. $\Gamma_{L,R}$ denote the
corresponding rate for the tunneling from (to) the left (right) reservoirs.
The penetration coefficient of the point-contact is $T$ for the empty
double-dot system or for the occupied second dot,
and it is $T'$ for the occupied first dot. The indices $m$ and $n$
denote the number of electrons penetrating to the right reservoirs at
time $t$.}
\label{mfig4}
\end{figure}

The currents in the double-dot system ($I_s$) and in the detector ($I_d$)
are given by Eqs.~(\ref{aam3}):
\begin{subequations}
\label{bbm5}
\begin{align}
&I_s=\sum_{m,n}m(\dot\sigma_{aa}^{m,n}+\dot\sigma_{bb}^{m,n}
+\dot\sigma_{cc}^{m,n})=\Gamma_R\sigma_{cc}
\label{bbm5a}\\
&I_d=\sum_{m,n}n(\dot\sigma_{aa}^{m,n}+\dot\sigma_{bb}^{m,n}
+\dot\sigma_{cc}^{m,n})\nonumber\\
&~~~~~~~~~~~~~~~~~~~~~~~~~~~~~~~~=D-(D-D')\sigma_{bb}
\label{bbm5b}
\end{align}
\end{subequations}
where $\sigma_{ij}=\sum_{m,n}\sigma_{ij}^{m,n}$. It follows from
Eq.~({\ref{bbm5b}) that the variation of the detector current
$\Delta I_d=I_d^{(0)}-I_d$ is given by Eq.~(\ref{am8}), where
$\bar\sigma =\sigma_{bb}$.

Performing summation in Eqs.~(\ref{bm5})
over the number of electrons arrived to the
collectors ($m,n$), we obtain the following Bloch-type equations for the
reduced density-matrix of the double-dot system:
\begin{subequations}
\label{bm6}
\begin{align}
&\dot\sigma_{aa}  = -\Gamma_L\sigma_{aa}+\Gamma_R\sigma_{cc}
\label{bm6a}\\
&\dot\sigma_{bb}  = \Gamma_L\sigma_{aa}+i\Omega_0 (\sigma_{bc}-\sigma_{cb})
\label{bm6b}\\
&\dot\sigma_{cc}  = -\Gamma_R\sigma_{cc}-i\Omega_0 (\sigma_{bc}-\sigma_{cb})
\label{bm6c}\\
&\dot\sigma_{bc}  = i\epsilon\sigma_{bc}+i\Omega_0 (\sigma_{bb}-\sigma_{cc})
\nonumber\\
&~~~~~~~~~~~~~~~~~~~~~~~~~~~~~~~~~~~
-\frac{1}{2}(\Gamma_R+\Gamma_d)\sigma_{bc},
\label{bm6d}
\end{align}
\end{subequations}
where $\Gamma_d$ is the dephasing rate generated by the detector,
Eq.~(\ref{cm5}). Solving these equations in the limit $t\to\infty$ we find
the following expression for the current $I_s$, Eq.~(\ref{bbm5a}),
flowing through the double-dot system
\begin{equation}
I_s=\frac{(\Gamma_R+\Gamma_d)\Omega_0^2}
{\epsilon^2+\frac{(\Gamma_R+\Gamma_d)^2}{4}
+\Omega_0^2(\Gamma_R+\Gamma_d)
\left (\frac{2}{\Gamma_R}+\frac{1}{\Gamma_L}\right )}
\label{bm8}
\end{equation}

By analyzing Eq.~(\ref{bm8}) one finds that the decoherence rate, $\Gamma_d$, Eq.~(\ref{cm5}), generated by the measurement, would affect the resonant current, $I_s$. As an example, we display it in Fig.~\ref{mfig5}
for three values of decoherence rate: $\Gamma_d=0$,  $\Gamma_d=4\Omega_0$ and $\Gamma_d=16\Omega_0$. We find that
for small $\epsilon$ the current decreases with $\Gamma_d$,
while for large $t$ the average distribution of an electron
in the dots remains the same.
However, for larger values of $\epsilon$ the current
{\em increases} with $\Gamma_d$.
It reflects electron delocalization in a double-well
system, Fig.~\ref{mfig3} b, due to continuous monitoring of the
charge in the left dot. Thus, by measure the resonant current $I_s$ for different values of $\epsilon$ or (and) $\Gamma_{L,R}$, one can extract the desirable decoherence rate, $\Gamma_d$

\begin{figure}[tbh]
\includegraphics[width=8cm]{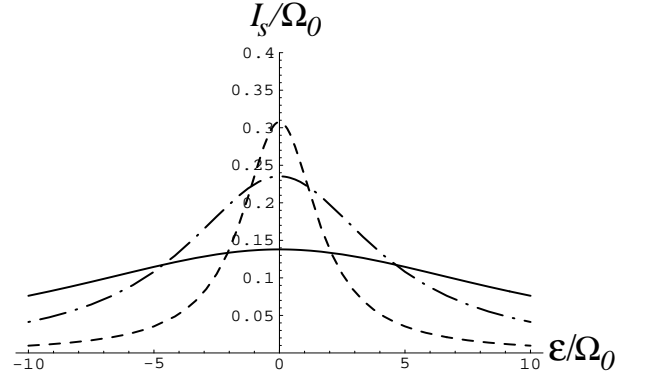}
\caption{ Electron current through the double-dot, Eq.~(\ref{bm8}),
for $\Gamma_L=\Gamma_R=\Omega_0$
as a function of the level displacement $\epsilon =E_2-E_1$.
The curves correspond to different values of the dephasing
rate: $\Gamma_d=0$ (dashed),  $\Gamma_d=4\Omega_0$ (dot-dashed) and
$\Gamma_d=16\Omega_0$ (solid).}
\label{mfig5}
\end{figure}

\section{Continuous measurement of decay to a non-Markovian reservoir}
 \subsection{Tunneling to continuum.}

Consider tunneling of a particle (electron) from a potential well (quantum dot) to a reservoir, Fig.~\ref{nfig1}. The system is described by the following Hamiltonian
\begin{align}
H=E_0|0\rangle\langle 0|+\sum_rE_r|r\rangle\langle r|+\sum_r\Omega_r(|r\rangle\langle 0|+|0\rangle\langle r|)
\label{h1}
\end{align}
Here $|0\rangle$ is a localized state in the well and $|r\rangle$ denotes  extended states of the reservoir.
\begin{figure}[h]
\includegraphics[width=5cm]{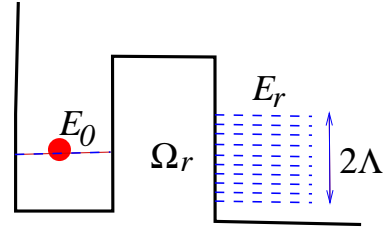}
\caption{Tunneling of a particle to continuum from a localized state inside the well to a reservoir of a finite bandwidth $\Lambda$.}
\label{nfig1}
\end{figure}

The tunneling of a particle to the reservoir is described by the Schr\"odiger equation,
\begin{align}
i\partial_t |\Psi (t)\rangle =H|\Psi (t)\rangle
\label{sch}
\end{align}
where $|\Psi (t)\rangle$ can be written as
\begin{align}
|\Psi (t)\rangle=b_0(t)|0\rangle +\sum_rb_r(t)|r\rangle
\label{na3}
\end{align}
Here $b_{0}(t)$ is probability amplitude for finding the particle at the state $|0\rangle$ inside the well and $b_{r}(t)$ is the same for the state $|r\rangle$ inside the reservoir. Substituting Eq.~(\ref{na3}) into Eq.~(\ref{sch}) and performing the Laplace transform, Eq.~(\ref{lap})
we can rewrite Eq.~(\ref{sch}) as
\begin{subequations}
\label{na4}
\begin{align}
&(E-E_0)\tilde b_0(E)-\sum_r\Omega_r\tilde b_r(E)= i\label{na4a}\\
&(E-E_r)\tilde b_r(E)-\Omega_r\tilde b_0(E)=0
\label{na4b}
\end{align}
\end{subequations}
where the right-hand-side corresponds to the initial conditions.

Solving Eqs.~(\ref{na4}) in the continuous limit,  $\sum_r\to\int\rho(E_r)dE_r$, where $\rho(E_r)$ is the density of state, we find
\begin{align}
\tilde b_0(E)={i\over E-E_0-\int\limits_{-\infty}^{\infty} {\Omega^2(E_r)\rho(E_r)\over E-E_r}dE_r}
\label{nm7}
\end{align}
with $\Omega_r^{}\equiv \Omega (E_r^{})$. In the case of Markovian reservoir (wide-band limit, $\Lambda\to\infty$), the density of states and the coupling $\Omega (E_r^{})$ are independent of $E_r^{}$. Then integration over $E_r$ in Eq.~(\ref{nm7}) can be easily performed, thus obtaining
\begin{align}
\tilde b_0(E)={i\over E-E_0+i{\Gamma\over2}}
\label{nm71}
\end{align}
where $\Gamma =2\pi\Omega^2\rho$. Note that for infinite reservoir, the density of states $\rho\sim L\to\infty$, where $L$ is the reservoir's size, but $\Omega^2\sim 1/L\to 0$, so the product $\Omega^2\rho$ (spectral density function) remains finite.

The amplitude $b_{0}(t)$ is obtained from $\tilde b_{0}(E)$ via the inverse Laplace transform,
\begin{align}
b_{0}(t)=\int\limits_{-\infty}^\infty \tilde b_{0}(E)e^{-iEt}{dE\over 2\pi}=e^{-iE_0t-{\Gamma\over2}t}
\label{ninvlap}
\end{align}
As a result, probability of finding a particle inside the well (survival probability) $P_0(t)=|b_0^{}(t)|^2=e^{-\Gamma t1}$.
Thus in the wide-band limit, the particle initially localized inside the quantum well, decays exponentially to the reservoir.

Consider now a (non-Markovian) reservoir of a finite band-width, Fig.~\ref{nfig1}. It corresponds to a periodic one-dimensional chain of $N$ quantum wells, with the nearest-neighbor coupling $\lambda$, shown in Fig.~\ref{figap1} and
describing  by the following Hamiltonian
\begin{figure}[h]
\includegraphics[width=8cm]{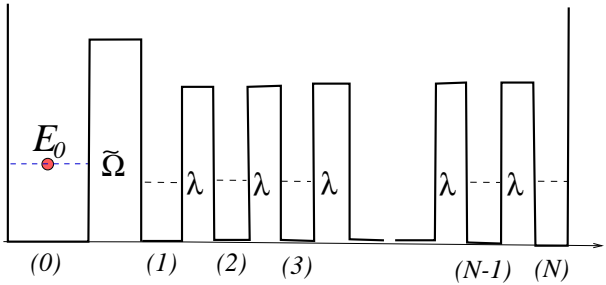}
\caption{Tunneling well to a periodic chain of quantum  wells with the nearest-neighbor coupling $\lambda$}
\label{figap1}
\end{figure}
\begin{align}
H_N=\sum_{n=1}^{N-1} \lambda(|n\rangle\langle n+1|+|n+1\rangle\langle n|)
\label{tbham1}
\end{align}
The state $|0\rangle$ of the quantum well is coupled with the first site of the chain by coupling $\tilde\Omega$, so the total Hamiltonian is $H=E_0|0\rangle\langle 0|+H_N+\tilde\Omega(|0\rangle\langle 1|+|1\rangle\langle 0|)$. By diagonalizing $H_N$ one arrives to Eq.~(\ref{h1}) with
\begin{subequations}
\label{tbham2}
\begin{align}
|r\rangle =&\sqrt{{2\over N+1}}\sum_{n=1}^N\sin \left({r\pi\over N+1}n\right)|n\rangle\label{tbham2a}\\
E_r^{}=&-2\lambda\cos \left({r\pi\over N+1}\right),~~{\rm for}~~ r=1,\ldots ,N\ ,
\label{tbham2b}
\end{align}
\end{subequations}
so that $-2\lambda<E_r<2\lambda$, and the corresponding spectral function is
\begin{align}
\Omega^2(E_r)\rho(E_r)={\Gamma\over2\pi}\sqrt{1-{E_r^2\over4\lambda^2}}
\label{tbham5}
\end{align}
where $\rho(E_r^{})=(dE_r/dr)^{-1}$ is the density of states and $\Gamma=\tilde\Omega^2/\lambda$. Here the band-center $E_R=0$.
The Markovian case (wide-band limit) corresponds to $\lambda\to\infty$. We assume that $\Gamma$ remains finite in this limit, which requires $\tilde\Omega\propto\sqrt{\lambda}\ll\lambda$.

In our calculations we approximate the spectral function (\ref{tbham5}) by the Lorentzian
\begin{align}
\Omega^2(E_r)\rho (E_r)={\Gamma\over 2\pi}{\Lambda^2\over (E_r-E_R)^2
+\Lambda^2}\, ,
\label{lor}
\end{align}
where $E_R$ is the Lorentzian center, and $\Lambda=2\sqrt{2}\lambda$, providing  the same curvature at the band-center that of Eq.(\ref{tbham5}). Such approximation allows us to treat the problem analytically without loosing its main physical features. For instance, Fig.~\ref{nfig5} shows survival probability $P_0(t)=|b_0(t)|^2$, obtained from Eqs.~(\ref{na4}) and  (\ref{tbham2}) for $N=250$, $\lambda =3\Gamma$ and $E_0=\Gamma$ (dashed line) in comparison with the Lorentzian spectral function, Eq.~(\ref{lor}), (solid line). One finds that both curves almost coincide. This confirms that the Lorentzian (\ref{lor}) is a very good approximation for finite band-width reservoirs, Eq.(\ref{tbham5}).
\begin{figure}[h]
\includegraphics[width=7cm]{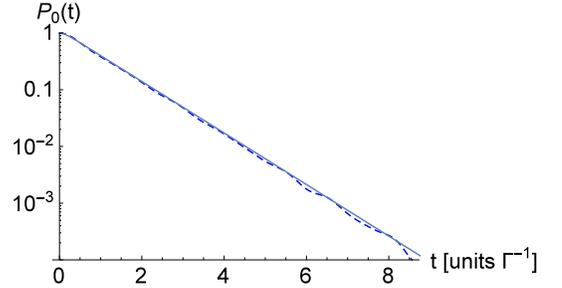}
\caption{Probability of finding the particle at its initial state at time $t$. Dashed line corresponds to periodic chain of $N=250$ coupled wells, and solid line shows continuous limit, $N\to\infty$, where the density of states is the Lorentzian, Eq.~(\ref{lor}).}
\label{nfig5}
\end{figure}

Substituting Eq.~(\ref{lor}) in Eq.~(\ref{nm7}), we can evaluate the integral by closing the integration contour into lower complex $E_r$-plane. As a result, Eq.~(\ref{nm7}) becomes
\begin{align}
(E-E_0)\tilde b_0(E)-{\Gamma\Lambda\over 2(E-E_R+i\Lambda) }\tilde b_0(E)=i
\label{na6}
\end{align}
(In the following we choose $E_0=0$.)
Using the inverse Laplace transform, Eq.~(\ref{ninvlap}), we obtain
\begin{align}
b_0^{}(t)=\, e^{-{Q\over2}t}
\Big[\cosh\Big({St\over2}\Big)
+{Q\over S}\sinh\Big({St\over2}\Big)\Big]
\label{nm9}
\end{align}
where $Q=\Lambda+iE_R$ and $S=\sqrt{Q^2-2\Lambda\Gamma}$. In the limit $\Lambda\to\infty$ we return to the Markovian case by reproducing the exponential decay, Eq.~(\ref{ninvlap}), whereas for finite $\Lambda$, Eq.~(\ref{nm9}) reproduces two-exponential decay. The difference with Markovian case is mostly significant for small times $t$. Indeed, the probability of decay to the Markovian reservoir, Eq.~(\ref{ninvlap}), reveals the irreversible dynamics, $1-|b_0(t)|^2_{}=\Gamma t+{\cal O}[t^2]$, whereas for the non-Markovian case, Eq.~(\ref{nm9}) the dynamics is reversible,
\begin{align}
1-P_0(t)={\Gamma\Lambda\over2}t^2+{\cal O}[t^3]
\end{align}
This makes a crucial importance for the quantum Zeno effect in continuous monitoring of decay to the non-Markovian reservoir. For a most effective treatment of this problem, we introduce a new basis for states of the non-Markovian reservoir.

\subsection{New basis of the reservoir's states.}

Consider Eq.~(\ref{na6}) for the amplitude $\tilde b_0(t)$. Let us introduce the  auxiliary amplitude (c.f. with Ref.~[\onlinecite{eg2}])
\begin{align}
\tilde b_R(E)={\bar\Omega\over E-E_R+i\Lambda}\tilde b_0(E)
\label{na7}
\end{align}
where
\begin{align}
\bar\Omega=\sqrt{{\Gamma\Lambda\over2}}
\label{ombar}
\end{align}
Then Eqs.~(\ref{na6}), (\ref{na7}) can be rewritten as
\begin{subequations}
\label{na8}
\begin{align}
&(E-E_0)\tilde b_0(E)-\bar\Omega \tilde b_R(E)=i\label{na8a}\\
&(E-E_R+i\Lambda )\tilde b_R(E)-\bar\Omega \tilde b_0(E)=0
\label{na8b}
\end{align}
\end{subequations}

Let us demonstrate that Eqs.~(\ref{na8}) describe the particle in a  double-well, shown in Fig.~\ref{nfig2}, where the second well is a fictitious one, coupled with a fictitious Markovian reservoir, with the coupling $\Omega$ and density of states $\rho$, such that  $\pi\Omega^2\rho=\Lambda$. This system is described by the Hamiltonian
\begin{align}
&H=E_0|0\rangle\langle 0|+E_R|R\rangle\langle R|
+\sum_{r'}E_{r'}|r'\rangle\langle r'|\nonumber\\
&+\bar\Omega(|R\rangle\langle 0|+|0\rangle\langle R|)
+\sum_{r'}\Omega(|r'\rangle\langle R|+|R\rangle\langle r'|)
\label{a1n}
\end{align}
Comparing (\ref{a1n}) with the original Hamiltonian, Eqs.~(\ref{h1}), we find that the reservoir's (extended) states $|r\rangle$ are split into the two components
\begin{align}
\sum_r|r\rangle\langle r|=|R\rangle\langle R|+\sum_{r'}|r'\rangle\langle r'|
\label{compn}
\end{align}
where $|r'\rangle$ represent the extended states of the fictitious {\em Markovian} reservoir.

\begin{figure}[h]
\includegraphics[width=6cm]{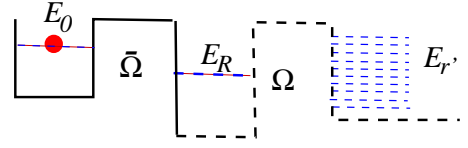}
\caption{Tunneling from the left well to continuum through the fictitious well. The level $E_R$ is at the Lorentzian center.}
\label{nfig2}
\end{figure}

Now the particle wave function can be written in this new basis as
\begin{align}
|\Psi(t)\rangle=b_0(t)|0\rangle +b_R(t)|R\rangle+\sum_{r'}b_{r'}(t)|r'\rangle
\label{na3p}
\end{align}
Substituting it in the Schr\"odinger equation
$i\partial_t|\Psi(t)\rangle=H|\Psi(t)\rangle$ we find
\begin{subequations}
\label{nac4}
\begin{align}
&i\dot b_0(t)=E_0b_0(t)+\bar\Omega\, b_R(t)\label{nac4a}\\
&i\dot b_R(t)=E_Rb_R(t)+\bar\Omega b_0(t)+\sum_{r'}\Omega\, b_{r'}(t)
\label{nac4b}\\
&i\dot b_{r'}(t)=E_{r'}b_{r'}(t)+\Omega\, b_R(t)
\label{nac4c}
\end{align}
\end{subequations}
Resolving Eq.~(\ref{nac4c}) and substituting it into Eq.~(\ref{nac4b}), we obtain in the continuous limit, $\sum_{r'}\to \int\rho dE_{r'}$,
\begin{subequations}
\label{nab8}
\begin{align}
&i\dot b_0(t)=E_0 b_0(t)+\bar\Omega b_R(t)\label{nab8a}\\
&i\dot b_R(t)=(E_R-i\Lambda )b_R(t)+\bar\Omega\, b_0(t)
\label{nab8b}
\end{align}
\end{subequations}
After the Laplace transform, these equations coincide with Eqs.~(\ref{na8}).

\subsection{Continuous monitoring with point-contact detector.}

Consider now the continuous monitoring of the decay to a non-Markovian reservoir of a finite band width, by placing the Point-Contact (PC) detector in close proximity to the quantum well, Fig.~\ref{nfig3}. Then the opening of PC decreases due repulsive electrostatic field of the electron, occupying the quantum dot. This results in increase of the barrier hight, and therefore in decrease on the electric current ($I$), flowing through the PC.  However, when the electron tunnels to the reservoir (to the fictitious well of Fig.~\ref{nfig2}), its electric field near the PC decreases and the corresponding electric current increases, $I\to I'$ in Fig.~\ref{nfig3}. Thus one can monitor the electron decay to continuum via the PC current.
\begin{figure}[h]
\includegraphics[width=6cm]{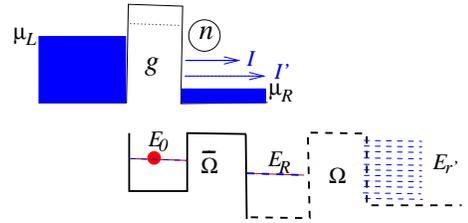}
\caption{Measurement of the quantum-well population with a point-contact detector. The current increases whenever the electron leaves the well. $n$ denotes the number of electrons arriving the right lead at time $t$.}
\label{nfig3}
\end{figure}

The entire system is described by the following Hamiltonian $H=H_0+H_{PC}+H_{int}$, where $H_0$ is given Eq.~(\ref{a1n}), $H_{PC}$ describes the PC detector, and $H_{int}$ is the interaction term. We use
\begin{eqnarray}
H_{PC}&=&\sum_l \bar E_l c_l^\dagger c_l+\sum_r \bar E_r c_r^\dagger c_r
+\sum_{l,r} [g\, c_l^\dagger c_r+H.c.]
\nonumber\\
H_{int}&=&\sum_{l,r} [\delta g\, c_l^\dagger c_r+H.c.]
|0\rangle\langle 0|\ ,
\label{bn1}
\end{eqnarray}
where the operators $c_{l(r)}^\dagger (c_{l(r)})$ corresponds to the creation
(annihilation) of electron in the state $\bar E_l(\bar E_r)$, belonging to the left (right) lead and $g$ is tunneling coupling
between these states. The quantity $\delta g=g'-g$
represents variation of the point contact hopping amplitude,
when the dot is occupied by the electron.

The entire system undergoes continues Schr\"odinger evolution, describing by the time-dependent Schr\"odinger $i\partial_t|\Psi (t)\rangle =H|\Psi (t)\rangle$, where $|\Psi (t)\rangle$ is the total many-particle wave-function including the PC detector. The initial condition, $|\Psi (0)\rangle$, corresponds to the occupied quantum dot when the leads are filled up to the Fermi levels $\mu_L$ and $\mu_R$, Fig.~\ref{nfig3}. The probability of finding the dot occupied at time time $t$ is $\sigma_{00}^{}(t)={\rm Tr}|\langle\Psi (t)|0\rangle|^2$, where the tracing takes place over all variables of the system. Solving the time-dependent Schr\"odinger equation one can evaluate $\sigma_{00}^{}(t)$.

The problem can be solved analytically in the large bias limit, $V_d=\mu_L-\mu_R$, by using a new basis of the reservoir's states, Eq.~(\ref{compn}). Then in the large bias limit the many-body Schr\"odinger  equation for $|\Psi (t)\rangle$ can be transformed  to master equations for the reduced density matrix
$\sigma_{jj'}^{(n)}(t)={\rm Tr}\langle j,n|\Psi (t)\rangle \langle\Psi (t)|j',n\rangle$, where $j(j')=\{ 0,R\}$, and $n$ is the number of electrons, arriving the right lead at time $t$, with $n/t$ is the average PC current. Using Eq.~(\ref{d1}), one finds (c.f. with Eqs.(44a)-(44c) of Ref.~[\onlinecite{eg2}])
\begin{subequations}
\label{cn}
\begin{align}
&\dot{\sigma}_{00}^{(n)} = -I\sigma_{00}^{(n)}+I\sigma_{00}^{(n-1)}
+i\bar\Omega (\sigma_{0R}^{(n)}-\sigma_{R0}^{(n)})
\label{cn1}\\
&\dot{\sigma}_{RR}^{(n)} =  -(I'+2\Lambda )\sigma_{RR}^{(n)}
+I'\sigma_{RR}^{(n-1)}\nonumber\\
&~~~~~~~~~~~~~~~~~~~~~~~~~~~~~~~~~~~~+i\bar\Omega (\sigma_{R0}^{(n)}-\sigma_{0RR}^{(n)})
\label{cn2}\\
&\dot{\sigma}_{0R}^{(n)} = i\epsilon\sigma_{0R}
+i\bar\Omega (\sigma_{00}^{(n)}-\sigma_{RR}^{(n)})
-\Big(\frac{I+I'}{2}+\Lambda\Big)\sigma_{0R}^{(n)}
\nonumber\\
&~~~~~~~~~~~~~~~~~~~~~~~~~~~~~~~~~~~~~~~~~
+\sqrt{I\, I'}\sigma_{0R}^{(n-1)}
\label{ncn3}
\end{align}
\end{subequations}
where $\bar\Omega=\sqrt{\Gamma\Lambda /2}$, Eq.~(\ref{ombar}) and $\epsilon =E_0-E_R$. Here $I=2\pi g^2\rho_L\rho_RV_d$ is a current though the PC, when the quantum dot is occupied. Respectively, $I'=2\pi{g}^{\prime\, 2}\rho_L\rho_RV_d$ is is the same for the empty dot.

The reduced density-matrix $\sigma_{jj'}^{(n)}(t)$ describes both the tunneling electron and the PC current. By tracing it over $n$, we find probability of the dot's occupation, $\sigma_{00}^{}(t)=\sum_n\sigma_{00}^{(n)}(t)$. Performing this procedure in Eqs.~(\ref{cn}) we obtain (c.f. with Eqs.(45a)-(45c) of Ref.~[\onlinecite{eg2}])
\begin{subequations}
\label{na11}
\begin{align}
&\dot\sigma_{00}=i\bar\Omega(\sigma_{0R}-\sigma_{R0})\label{a11a}\\
&\dot\sigma_{RR}=i\bar\Omega(\sigma_{R0}-\sigma_{0R})-2\Lambda
\sigma_{RR}\label{na11b}
\\
&\dot\sigma_{0R}=i\epsilon\sigma_{0R}+
i\bar\Omega(\sigma_{00}-\sigma_{RR})-\left({\Gamma_d\over2}+\Lambda\right)
\sigma_{0R}
\label{na11c}
\end{align}
\end{subequations}
where $\sigma_{jj'}^{}(t)=\sum_n\sigma_{jj'}^{(n)}(t)$ and $\Gamma_d=(\sqrt{I}-\sqrt{I'})^2$, Eq.~(\ref{cm5}).
These equations are of the Lindbladt (Bloch)-type Master equations and have a  clear physical meaning.  Indeed, in the case of no interaction with the PC detector ($I=I'$ and therefore $\Gamma_d=0$), one easily obtains Eqs.~(\ref{na11}) directly from Eqs.~(\ref{nab8}), taking into account that $\sigma_{jj'}^{}(t)=b_j^{}(t)b_{j'}^*(t)$. Hence, the interaction with the PC detector generates an additional damping (decoherence) rate ($\Gamma_d$) in Eq.~(\ref{na11c}) for the off-diagonal density-matrix element, $\sigma_{0R}(t)$.

It order to solve Eqs.~(\ref{na11}) it is useful to apply Laplace transform, $\sigma (t)\to\tsigma(E)$, Eq.~(\ref{lap}), thus obtaining
\begin{subequations}
\label{na13}
\begin{align}
&E\tsigma_{00}+\bar\Omega(\tsigma_{0R}-\tsigma_{R0})=i\label{na13a}\\
&(E+2i\Lambda )\tsigma_{RR}+\bar\Omega(\tsigma_{R0}-\tsigma_{0R})=0\label{na13b}
\\
&\left[E+\Delta+i\Lambda\left(1+{\Gamma_d\over 2\Lambda}\right)\right]\tsigma_{0R}+
\bar\Omega(\tsigma_{00}-\tsigma_{RR})=0
\label{na13c}
\end{align}
\end{subequations}
Let us analyze Eqs.~(\ref{na13}) in the limit of large decoherence, $\Gamma_d\to\infty$, corresponding to ``strong'' measurement. In parallel, we also take the limit of $\Lambda\to\infty$, while their ratio, $x=\Lambda/\Gamma_d$ is keeping constant. Solving Eqs.~(\ref{na13}) in this limit, we find
\begin{align}
\tsigma_{00} (E)=\frac{i \left(1+2x\right)}{E\left(1+2x\right)+2\,i\Gamma x}
\label{na17}
\end{align}

Performing the inverse Laplace transform, Eq.~(\ref{ninvlap}),
by closing the contour of integration over the pole of $\tsigma_{00} (E)$, we finally obtain
\begin{align}
\sigma_{00}^{}(t)=\exp\Big(-{2\,x\Gamma t\over 1+2x}\Big)
\label{scaling}
\end{align}
This result is very remarkable, since it displays the influence of continuous measurement for the both Markovian and non-Markovian environments. Indeed, for  Markovian reservoir, $\Lambda\to\infty$ and respectively, $x\to\infty$. As a result, $\sigma_{00}^{}(t)=\exp (-\Gamma t)$. It implies that there is no measurement effect on decay to Markovian reservoir. However, in the opposite case, of very strong continuous measurement, $\Gamma_d\to\infty$, and $\Lambda$ is finite, corresponding to $x=0$,, one finds from Eq.~(\ref{scaling}) that $\sigma_{00}^{}(t)=1$. This implies no decay to continuum (quantum Zeno effect).

On the first sight, the Zeno effect looks very paradoxical, since large decorerence rate implies large energy fluctuation of the system due to interaction with the PC detector. One could expect that these fluctuations would destabilize the system, instead of freezing, as predicted by Eq.~(\ref{scaling}). However, it is not the case \cite{xq2}. Indeed, for a finite reservoir band, there are no available  reservoir states with such large energies. As a result, the system cannot decay to continuum, despite large energy fluctuation, generated by the detector.
On the other hand, the spectral density function of Markovian reservoir is constant. Therefore, the energy fluctuations of the dot level are irrelevant for transition rates. As a result, one expects no Zeno effect at all, as predicted by Eq.~(\ref{scaling}) in the Markoviam limit, $x\to\infty$.

\section{Concluding remarks}

In this paper we investigated a reduction of the many-body Schr\"odinger equation to the Lindblad-type Master equation for quantum transport, with no assumption on weak coupling to the environment. As a result of such reduction we arrived to the particle-number resolved Master equations. Here we mainly concentrated on derivation and validity of the resulting Master equations. In particular, we demonstrated that the Markovian Master equations can be valid only in the high bias limit (strong non-equilibrium condition). Otherwise, such equations would not be Markovian.

A possible application of the wave-function approach for treatment of quantum transport in not restricted by the Master equations, discussed in this paper. For instance, one can use a different basis for the many-body wave-function, as in a recently proposed single-electron approach \cite{g0}. The latter based on the single-electron Ansatz for the many-electron wave-function, yields simple expressions for the tunneling currents in the presence of fluctuating environment \cite{gae}. In contrast with the Master equation approach, the single-electron approach is applicable to any bias, including linear response. A possible combination of the single-electron approach with that discussed in the present paper is a topic of current research. It could allow us to obtain non-Markovian Master equations valid for small bias. To achieve this goal,  would be very important for different applications and for better understanding of quantum-classical transition.

\begin{acknowledgements}

The author is very grateful to his collaborators in the original papers
constitute this review, in particular to Yakov Prager, Brahim Elattari and Xin-Qi Li.
\end{acknowledgements}

\end{document}